\documentclass[12pt,onecolumn,letter]{IEEEtran}

\usepackage[pdftex]{graphicx}
\DeclareGraphicsExtensions{.png,.pdf,.jpg,.eps}

\usepackage{amsmath}
\usepackage{amsfonts,amssymb}

\usepackage{latexsym}
\usepackage{amssymb}
\usepackage{amsmath}
\usepackage{multirow}
\usepackage{footnote}
\usepackage{color}
\usepackage{bm}

\newcommand{\Tr}{{\rm Tr}\,}
\def\mix{\mathop{\rm mix}}
\def\argmin{\mathop{\rm argmin}}

\def\bZ{\mathbb Z}
\def\bC{\mathbb C}
\def\bF{\mathbb F}
\def\Ker{\mathop{\rm Ker}}
\def\rE{{\rm E}}
\def\Pr{{\rm Pr}}

\providecommand{\url}[1]{#1}
\csname url@samestyle\endcsname

\providecommand{\BIBforeignlanguage}[2]{{%
\expandafter\ifx\csname l@#1\endcsname\relax
\typeout{** WARNING: IEEEtranS.bst: No hyphenation pattern has been}%
\typeout{** loaded for the language `#1'. Using the pattern for}%
\typeout{** the default language instead.}%
\else
\language=\csname l@#1\endcsname
\fi
#2}}
\providecommand{\BIBdecl}{\relax}
\BIBdecl

\def\Label#1{\label{#1}\ \text{[\ #1\ ]}\ }
\def\Label#1{\label{#1}}

\begin{document}

\title{Finite-Block-Length Analysis in Classical and Quantum Information Theory
(Review paper)
}

\author{
Masahito Hayashi
\thanks{
M. Hayashi is with Graduate School of Mathematics, Nagoya University, 
Furocho, Chikusaku, Nagoya, 464-8602, Japan, and
Centre for Quantum Technologies, National University of Singapore, 3 Science Drive 2, Singapore 117542.
(e-mail: masahito@math.nagoya-u.ac.jp)
}}
\date{}
\maketitle

\begin{abstract}
Coding technology is used in several information processing tasks. In particular, when noise during transmission disturbs communications, coding technology is employed to protect the information. However, there are two types of coding technology: coding in classical information theory and coding in quantum information theory. Although the physical media used to transmit information ultimately obey quantum mechanics, we need to choose the type of coding depending on the kind of information device, classical or quantum, that is being used. In both branches of information theory, there are many elegant theoretical results under the ideal assumption that an infinitely large system is available. In a realistic situation, we need to account for finite size effects. The present paper reviews finite size effects in classical and quantum information theory with respect to various topics, including applied aspects.
\end{abstract}


\section{Introduction}
A fundamental problem in information processing is to transmit a message correctly via a noisy channel,
where the noisy channel is mathematically described by a probabilistic relation between input and output symbols.
To address this problem, we employ channel coding, which is composed of two parts: 
an encoder and a decoder.
The key point of this technology is the addition of redundancy to the original message to protect it from corruption by the noise.
The simplest channel coding is transmitting the same information three times as shown in Fig. \ref{F1}.
That is, when we need to send one bit of information, $0$ or $1$, we transmit three bits, $0,0,0$ or $1,1,1$.
When an error occurs in only one of the three bits,
we can easily recover the original bit.
The conversion from $0$ or $1$ to $0,0,0$ or $1,1,1$ is called an encoder and
the conversion from the noisy three bits to the original one bit is called a decoder.
A pair of an encoder and a decoder is called a code.

In this example, the code has a large redundancy and the range of correctable errors is limited. For example, if two bits are flipped during the transmission, we cannot recover the original message.
For practical use, we need to improve on this code,
that is, decrease the amount of redundancy and enlarge the range of correctable errors.

\begin{figure}[htbp]
\begin{center}
\includegraphics[scale=0.7]{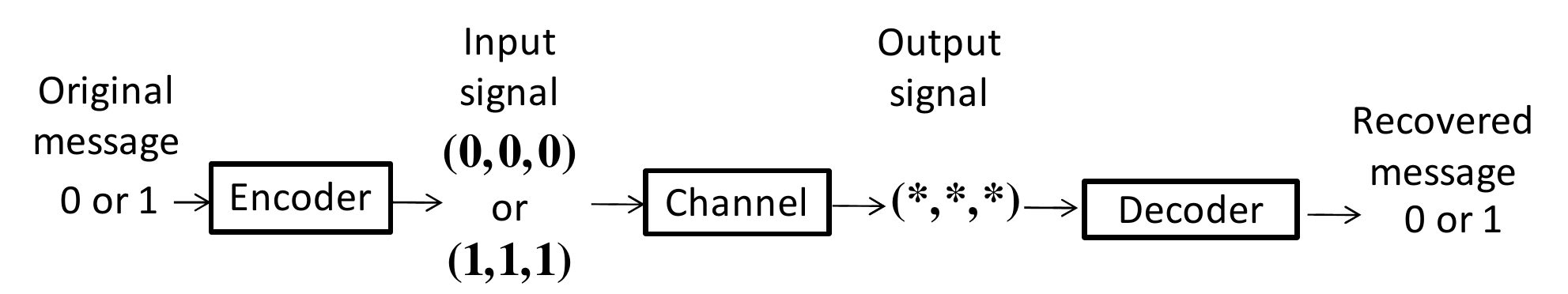}
\end{center}
\caption{Channel coding with three-bit code}
\Label{F1}
\end{figure}%

The reason for the large redundancy in the simple code described above is that the block-length (the number of bits in one block) of the code is only $3$.
In 1948, Shannon \cite{Shannon48} discovered that increasing the block-length $n$ can improve the redundancy and the range of correctable errors.
In particular, he clarified the minimum redundancy required to correct an error with probability almost $1$
with an infinitely large block-length $n$.
To discuss this problem, for a probability distribution $P$,
he introduced the quantity $H(P)$, which is called the (Shannon) entropy and expresses the uncertainty of the probability distribution $P$.
He showed that we can recover the original message by a suitable code
when the noise of each bit is independently generated subject to the probability distribution $P$, the rate of redundancy is the entropy $H(P)$,
and the block-length $n$ is infinitely large.
This fact is called the channel coding theorem.
Under these conditions, the limit of the minimum error probability
depends only on whether the rate of the redundancy is larger than the entropy $H(P)$ or not.

We can consider a similar problem when the channel is given as additive white Gaussian noise. 
In this case, we cannot use the term redundancy because its meaning is not clear.
In the following, instead of this term, we employ the transmission rate,
which expresses the number of transmitted bits per one use of the channel, to characterize the speed of the transmission.
In the case of an additive white Gaussian channel, the channel coding theorem is that
the optimal transmission rate is $ \frac{1}{2}\log (1+\frac{S}{N})$, where $\frac{S}{N}$ is the signal-noise ratio \cite[Theorem 7.4.4]{Gallager}.
However, we cannot directly apply the channel coding theorem to actual information transmission
because this theorem guarantees only the existence of a code with the above ideal performance.
To construct a practical code, we need another type of theory, which is often called coding theory.
Many practical codes have been proposed, depending on the strength of the noise in the channel, and have been used in real communication systems.
However, although these codes realize a sufficiently small error probability, no code could attain the optimal transmission rate.
Since the 1990s, turbo codes and low-density parity check (LDPC) codes have been actively studied as useful codes \cite{BGT,MN96}.
It was theoretically shown that they can attain the optimal transmission rate 
when the block-length $n$ goes to infinity.
However, still no actually constructed code could attain the optimal transmission rate.
Hence, many researchers have doubted what the real optimal transmission rate is.
Here, we should emphasize that any actually constructed code has a finite block-length and 
will not necessarily attain the conventional asymptotic transmission rate. 

On the other hand, in 1962, Strassen \cite{strassen} addressed this problem
by discussing the coefficient with the order $\frac{1}{\sqrt{n}}$ of the transmission rate,
which is called the second-order asymptotic theory.
The calculation of the second-order coefficient approximately gives the solution of the above problem, that is,
the real optimal transmission rate with finite block-length $n$.
Although he derived the second-order coefficient for the discrete channel,
he could not derive it for the additive white Gaussian channel.
Also, in spite of the importance of his result, many researchers overlooked his result
because his paper was written in German.
Therefore, the successive researchers had to recover his result 
without use of his derivation.
The present paper explains how this problem has been resolved even for additive white Gaussian channel
by tracing the long history of classical and quantum information theory.

Currently, finite block-length theory is one of hottest topics in information theory
and is discussed more precisely for various situations elsewhere \cite{Pol,Pol2,Hay2,Hay1,PPV2,PPV3,TK,SKT15,KV,TT13,Han5,YHN}.
Interestingly, in the study of finite-block-length theory,
the formulation of quantum information theory becomes closer to that of classical information theory \cite{Haya20}.

In addition to reliable information transmission, 
information theory studies data compression (source coding) and (secure) uniform random number generation.
In these problems, we address a code with block-length $n$.
When the information source is subject to the distribution $P$ and
the block-length $n$ is infinitely large, the optimal conversion rate is $H(P)$ in both problems.
Finite-length analysis also plays an important role in secure information transmission.
Typical secure information transmission methods are quantum cryptography and
physical layer security.
The aim of this paper is to review the finite-length analysis in these various topics in information theory.
Further, 
finite-length analysis has been developed in conjunction with an unexpected effect from 
the theory of quantum information transmission, which is often called quantum information theory.
Hence, we explain the relation between the finite-length analysis and quantum information theory.

The remained of the paper is organized as follows.
First, Section \ref{S15} outlines the notation used in information theory.
Then, Section \ref{S2} explains how the quantum situation is formulated as a preparation for later sections.
Section \ref{S3} reviews the idea of an information spectrum, which is a general method used in information theory.
The information spectrum plays an important role for developing the finite-length analysis later.
Section \ref{S4} discusses folklore source coding, which is the first application of finite-length analysis.
Then, Section \ref{S5} addresses quantum cryptography, which is the first application to an implementable communication system.
After a discussion of quantum cryptography,
Section \ref{S6} deals with second-order channel coding,
which gives a fundamental bound for finite-length of codes.
Finally, Section \ref{S7} discusses the relation between finite-length analysis and physical layer security.

\section{Basics of information theory}\Label{S15}
As a preparation for the following discussion, we provide the minimum mathematical basis for a discussion of information theory.
To describe the uncertainty of a random variable $X$ subject to the distribution $P_X$ on a finite set ${\cal X}$,
Shannon introduced the Shannon entropy 
$H(P_X):=- \sum_{x \in {\cal X}} P_X(x)\log P_X(x)$, which is often written as $H(X)$.
When $-\log P_X(x)$ is regarded as a random variable,
$H(P_X)$ can be regarded as its expectation under the distribution $P_X$.
When two distributions $P$ and $Q$ are given
the entropy is concave, that is, 
$\lambda H(P) +(1-\lambda )H(Q) \le H(\lambda P +(1-\lambda) Q) $ for $ 0<\lambda <1$.
Due to the concavity, the maximum of the entropy is $\log |{\cal X}|$, where 
$|{\cal X}|$ is the size of ${\cal X}$.
To discuss the channel coding theorem, we need to consider the 
conditional distribution $P_{Y|X}(y|x)=P_{Y|X=x}(y)$
where $Y$ is a random variable in the finite set ${\cal Y}$, 
which describes the channel with input system ${\cal X}$ and output system ${\cal Y}$.
In other words, the distribution of the value of the random variable $Y$ depends on the value of the random variable $X$.
In this case, we have the entropy $H(P_{Y|X=x})$ dependent on the input symbol $x \in {\cal X}$.

Now, we fix a distribution $P_X$ on the input system ${\cal X}$,
taking the average of the entropy $H(P_{Y|X=x})$,
we obtain the conditional entropy $\sum_{x \in {\cal X}} P_X(x) H(P_{Y|X=x})$, which is often written as $H(Y|X)$.
That is, the conditional entropy $H(Y|X)$ can be regarded as the uncertainty of the system ${\cal Y}$ when we know the value on ${\cal X}$.
On the other hand,
when we do not know the value on ${\cal X}$,
the distribution $P_Y$ on ${\cal Y}$ is given as 
$P_Y(y):=\sum_{x \in {\cal X}} P_X(x) P_{Y|X=x}(y)$.
Then, the uncertainty of the system ${\cal Y}$ 
is given as the entropy $H(Y):=H(P_Y)$, which is larger than the conditional entropy $ H(Y|X)$
due to the concavity of the entropy.
So, the difference $H(Y)-H(Y|X)$ can be regarded as the amount of knowledge in the system ${\cal Y}$ when we know the value on the system ${\cal X}$.
Hence, this value is called the mutual information between the two random variables $X$ and $Y$,
and is usually written as $I(X;Y)$.
Here, however,
we denote it by $I(P_X,P_{Y|X})$ to emphasize the dependence on the distribution $P_X$ over the
input system ${\cal X}$.

In channel coding, we usually employ the same channel $P_{Y|X}$ repetitively and independently ($n$ times).
The whole channel is written as
the conditional distribution 
$$
P_{Y^n|X^n=x^n}(y^n):=P_{Y|X=x_1}(y_1)\cdots P_{Y|X=x_n}(y_n)\;,
$$
where $x^n=(x_1, \ldots, x_n) \in {\cal X}^n$ and $y^n=(y_1, \ldots, y_n) \in {\cal Y}^n$.
This condition is called the memoryless condition.
In information theory, information intended to be sent to a receiver is called a message, and is distinguished from other types of information.
We consider the case that the sender sends a message, which is one element of the set 
${\cal M}_n:= \{1, \ldots, M_n\}$, where $M_n$ expresses the number of elements in the set.
Then, the encoder $E_n$ is written as a map from ${\cal M}_n $ to ${\cal X}^n$,
and the decoder $D_n$ is written as a map from ${\cal Y}^n$ to ${\cal M}_n $.
The pair of the encoder $E_n$ and the decoder $D_n$ is called a code.

Under this formulation, we focus on the decoding error probability 
$\epsilon(E_n,D_n):=\frac{1}{M_n}\sum_{m=1}^{M_n} 
(1- \sum_{y^n: D_n(y^n)=E_n(m)} P_{Y^n|X^n=E_n(m)}(y^n))$, which expresses the performance of a code 
$(E_n,D_n)$.
As another measure of the performance of a code $(E_n,D_n)$,
we focus on the size $M_n$, which is denoted by $|(E_n,D_n)|$ later.
Now, we impose the condition $\epsilon(E_n,D_n)\le \epsilon $ on our code $(E_n,D_n)$,
and maximize the size $|(E_n,D_n)|$.
That is, we focus on $M(\epsilon| P_{Y^n|X^n}):= \max_{(E_n,D_n)}\{|(E_n,D_n)|\; | \epsilon(E_n,D_n)\le \epsilon\} $.
In this context, the quantity $\frac{1}{n}\log M(\epsilon| P_{Y^n|X^n})$
expresses the maximum transmission rate under the above conditions.
The channel coding theorem characterizes the maximum transmission rate as follows.
\begin{align}
\lim_{n \to \infty}
\frac{1}{n}\log M(\epsilon| P_{Y^n|X^n}) =\max_{P_X} I(P_X,P_{Y|X}), \quad
0<\epsilon <1 .\Label{9-1-11}
\end{align}
The maximum value of the mutual information is called the capacity.

To characterize the mutual information, we introduce the relative entropy between two distributions $P$ and $Q$ as
$D(P\|Q):=\sum_{x \in {\cal X}}P(x) \log \frac{P(x)}{Q(x)}$.
When we introduce the joint distribution $P_{XY}(x,y):=P_X(x) P_{Y|X}(y|x)$
and the product distribution $(P_{X}\times P_Y)(x,y):= P_{X}(x) P_Y(y)$,
the mutual information is characterized as \cite{Shannon48,Gallager}
\begin{align}
I(P_X,P_{Y|X})= D(P_{XY} \|P_{X}\times P_Y)
=\min_{Q_Y}D(P_{XY} \|P_{X}\times Q_Y )
=\min_{Q_Y}\sum_{x}P_X(x) D(P_{Y|X=x}\| Q_Y ).
\end{align}
That is, the capacity is given as
\begin{align}
\max_{P_X} I(P_X,P_{Y|X})
&=\max_{P_X} D(P_{X}\times P_Y \| P_{XY})
=\max_{P_X} \min_{Q_Y}D(P_{X}\times Q_Y \| P_{XY}) \\
&=\max_{P_X} \min_{Q_Y}\sum_{x}P_X(x) D(P_{Y|X=x}\| Q_Y )
= \min_{Q_Y}\max_{P_X} \sum_{x}P_X(x) D(P_{Y|X=x}\| Q_Y ).
\end{align}
The final equation can be shown by using the mini-max theorem.

On the other hand, 
it is known that the relative entropy $D(P\|Q)$
characterizes the performance of statistical hypothesis testing 
when both hypotheses are given as distributions $P$ and $Q$.
Hence, we can expect an interesting relation between 
channel coding and statistical hypothesis testing.

As a typical channel, we focus on an additive channel.
When the input and output systems ${\cal X}$ and ${\cal Y}$ are given as 
the module $\bZ/d\bZ $,
given the input $X \in \bZ/d\bZ $, 
the output $Y \in \bZ/d\bZ $ is given as
$Y=X +Z$, where $Z$ is the random variable describing the noise and is subject to the distribution $P_Z$ on $\bZ/d\bZ$. 
Such a channel is called an additive channel or an additive noise channel.
In this case, 
the conditional entropy $H(Y|X)$ is $H(P_Z)$, because 
the entropy $H(P_{Y|X=x})$ equals $H(P_Z)$ for any input $x \in {\cal X}$,
and the mutual information $I(P_X,P_{Y|X})$ is given by $H(P_Y) - H(P_Z)$.
When the input distribution $ P_X$ is the uniform distribution, 
the output distribution $ P_Y$ is the uniform distribution and achieves the maximum entropy
$\log d$.
So, the maximum mutual information $\max_{P_X} I(P_X,P_{Y|X})$ is given as $\log d - H(P_Z)$.
That is, the maximum transmission equals $\log d - H(P_Z)$.
If we do not employ the coding, the transmission rate is $\log d$.
Hence, the entropy $H(P_Z)$ can be regarded as the loss of the transmission rate due to the coding.
In this coding, we essentially add the redundancy $H(P_Z)$ in the encoding stage.

It is helpful to explain concrete constructions of codes with the case of $d=2$, in which $\bZ/2 \bZ$ becomes the finite field $\bF_2$, which is the set $\{0,1\}$ with the operations of modular addition and multiplication,
when the additive noise $Z^n=(Z_1, \ldots, Z_n)$ is subject to the $n$-fold distribution $P_{Z}^n$ of $n$ independent and identical distributed copies of $Z \sim P_Z$. (From now on, we call such distributions ``iid distributions'' for short.) The possible transmissions are then elements of $\bF_2^n$ which is the set of $n$-dimensional vectors whose entries are either $0$ or $1$. 
In this case, we can consider the inner product 
in the vector space $\bF_2^n$ using the multiplicative and additive operations of $\bF_2$.
When $P_Z(1)=p$, the entropy $H(P_Z)$ is written as $h(p)$, where
the binary entropy is defined as
$h(p):= - p \log p - (1-p)\log (1-p)$. 
Since ${\cal X}^n=\bF_2^n$,
we choose a subspace $C$ of $\bF_2^n$ with respect to addition and we identify the message set ${\cal M}_n$ with $C$.
The encoder is given as a natural imbedding of $C$.
To find a suitable decoder, 
for a given element $[y]$ of the coset $\bF_2^n/C $,
we seek the most probable element $\Gamma([y])$ among $x+C$.
Hence, when we receive $y \in \bF_2^n$,
we decode it to $y- \Gamma([y])$.
It is typical to employ this kind of decoder.
To identify the subspace $C$, 
we often employ a parity check matrix $K$, in which, 
the subspace $C$ is given as the kernel of $K$.
Using the parity check matrix $K$,
the element of the coset $\bF_2^n/C $ can be identified using 
the image of the parity check matrix $K$, which is called the syndrome.
In this case, we denote the encoder by $E_K$.

Alternatively, when $\Gamma([y]) $ realizes $\max_{x^n \in [y]} P_Z^n(x^n)$,
the decoder is called the maximum likelihood decoder.
This decoder also gives the minimum decoding error $\epsilon(E_T,D)$.
As another decoder, we can choose $\Gamma([y]) $ such that
$\Gamma([y]) $ realizes $\max_{x^n \in [y]} |x^n|$,
where $|x^n|$ is the number of appearances of $1$ among $n$ entries.
This decoder is called the minimum distance decoder.
When $P_Z(0)>P_Z(1)$, the maximum likelihood decoder is the same as the minimum distance decoder. 
We denote the minimum distance decoder by $D_{K,\min}$.
This type of code is often called an error correcting code.

When most of the entries of the parity check matrix $K$ are zero, 
the parity check matrix $K$ is called an LDPC  matrix.
When the subspace $C$ is given as the kernel of 
an LDPC matrix,
the code is called the LDPC code.
In this case, it is known that 
a good decoder can be realized with a small calculation complexity
\cite{BGT,MN96}.
Hence, an LDPC code is used for practical purposes.

\section{Information Transmission via Quantum Coding}\Label{S2}
To discuss the information transmission problem, 
we eventually need to address the properties of the physical media carrying the information.
When we approach the ultimate limit of the information transmission rate as a theoretical problem,
we need to consider the case when individual particles express each bit of information. 
That is, we focus on the information transmission rate under such an extreme situation.
To realize the ultimate transmission rate, we need to use every photon (or every pulse) to describe one piece of information. 
Since the physical medium used to transmit the information behaves quantum mechanically under such conditions, 
the description of the information system needs to reflect this quantum nature.

Several researchers, such as Takahasi \cite{Takahasi}, started to consider the limit of optical communication in the 1960s.
In 1967, Helstrom \cite{Hel:1,Hel:2} started to systematically formulate this problem as a new type of information processing system based on quantum theory instead of 
an information transmission system based on classical mechanical input and output, which obeys conventional probability theory.
The study of information transmission based on such quantum media 
is called quantum information theory.
In particular, research on channel coding for quantum media is called quantum channel coding.
In contrast, information theory based on the conventional probability theory 
is called classical information theory 
when we need to distinguish it from quantum information theory,
even when the devices employ quantum effects in their insides,
because the input and the output are based on classical mechanics.
Quantum information theory in its earlier stage has been studied more deeply by Holevo
and is systematically summarized in his book \cite{HolP} in 1980.

\begin{figure}[htbp]
\begin{center}
\scalebox{0.7}{\includegraphics[scale=1]{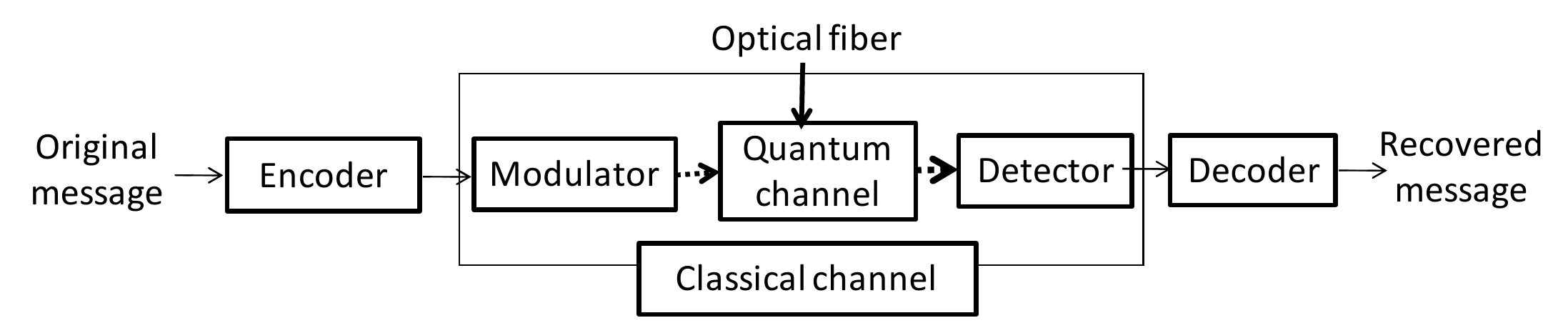}}
\end{center}
\caption{Classical channel coding for optical communication. 
Dashed thick arrows indicate quantum state transmission.
Normal thin arrows indicate classical information.}
\Label{F2}
\end{figure}%

\begin{figure}[htbp]
\begin{center}
\scalebox{0.7}{\includegraphics[scale=1]{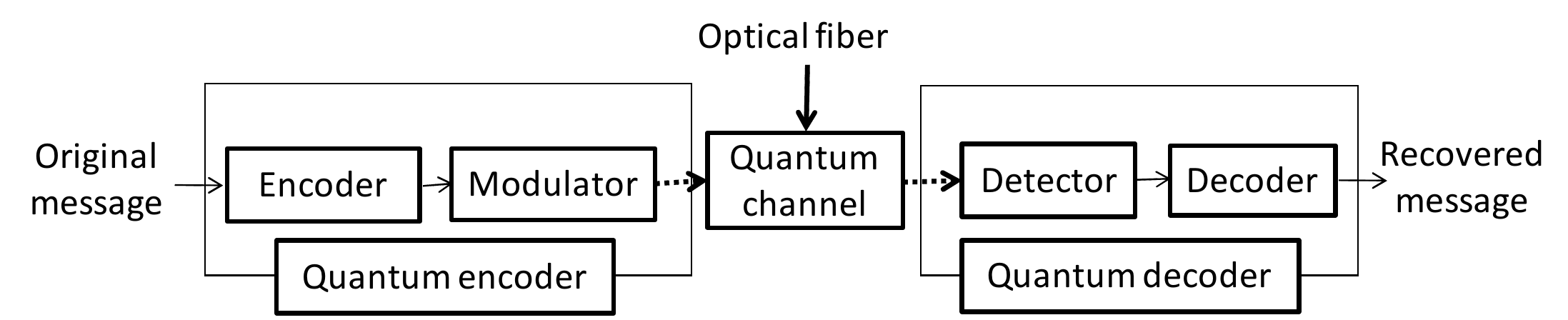}}
\end{center}
\caption{Quantum channel coding for optical communication.
Dashed thick arrows indicate quantum state transmission.
Normal thin arrows indicate classical information.}
\Label{F3}
\end{figure}%

Here, we point out that current optical communication systems are treated in the framework of classical information theory.
However, optical communication can be treated in both 
classical and quantum information theory as follows (Figs. \ref{F2} and \ref{F3}).
Because the framework of classical information theory cannot deal with a quantum system,
to consider optical communication within classical information theory,
we need to fix the modulator converting the input signal to the input quantum state
and the detector converting the output quantum state to the outcome, as shown in Fig. \ref{F2}.
Once we fix these,
we have the conditional distribution connecting the input and output symbols,
which describes the channel in the framework of classical information theory.
That is, we can apply classical information theory to the classical channel.
The encoder is the process converting the message (to be sent) to the input signal,
and the decoder is the process recovering the message from the outcome.

On the other hand, when we discuss optical communication within the framework of quantum information theory as shown in Fig. \ref{F3},
we focus on the quantum channel, whose input and output are given as quantum states.
When the quantum system is characterized by the Hilbert space ${\cal H}$,
a quantum state is given as a density matrix $\rho$ on ${\cal H}$, which is a positive-semi definite matrix with trace $1$.
Within this framework, we combine a classical encoder and a modulator into a quantum encoder, in which
the message is directly converted to the input quantum state.
Similarly, we combine 
a classical encoder and a detector into a quantum decoder, in which
the message is directly recovered from the output quantum state.
Once the optical communication is treated in the framework of quantum information theory,
our coding operation is given as the combination of a quantum encoder and a quantum decoder.
This framework allows us to employ physical processes across multiple pulses as
a quantum encoder or decoder,
so quantum information theory clarifies how much such a correlating operation enhances the
information transmission speed.
It is also possible to fix only the modulator and discuss 
the combination of a classical encoder and a quantum decoder,
which is called classical-quantum channel coding, as shown in Fig. \ref{F4}.
A classical-quantum channel is given as a map from 
an element $x$ of the input classical system
${\cal X}$ to an output quantum state $\rho_x$,
which is given as a density matrix on the output quantum system ${\cal H}$. 

\begin{figure}[htbp]
\begin{center}
\scalebox{0.7}{\includegraphics[scale=1]{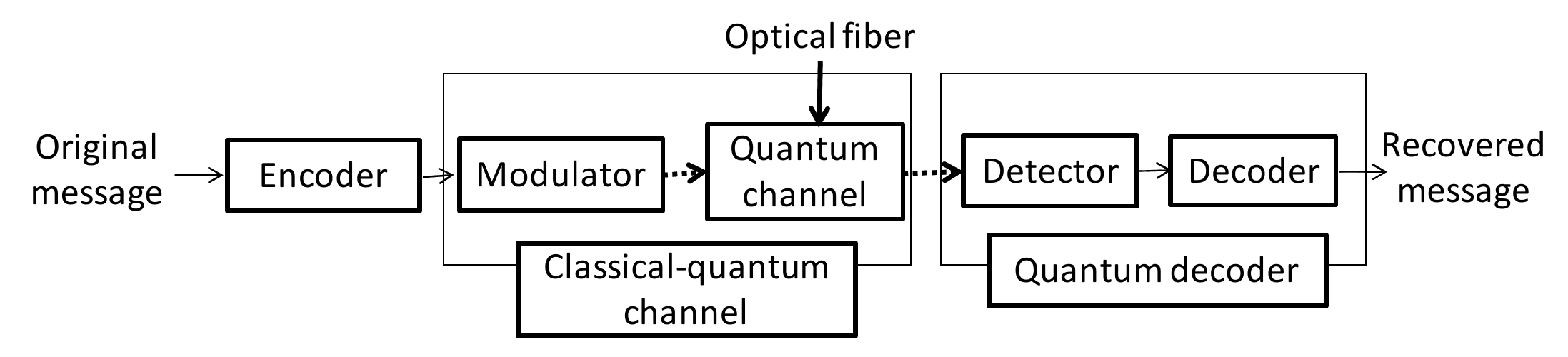}}
\end{center}
\caption{Classical-quantum channel coding for optical communication.
Dashed thick arrows indicate quantum state transmission.
Normal thin arrows indicate classical information.}
\Label{F4}
\end{figure}%

Here, we remark that the framework of quantum information theory mathematically 
contains the framework of classical information theory as the commutative special case, that is, the case when all $\rho_x$ commute with each other. 
This character is in contrast to 
the fact that a quantum Turing machine does not contain the conventional Turing machine as the commutative special case.
Hence, when we obtain a novel result in quantum information theory
and it is still novel even in the commutative special case,
it is automatically novel in classical information theory.
This is a major advantage and became a driving force for later 
unexpected theoretical developments.

A remarkable achievement of the early stage was made by Holevo in 1979, who obtained a partial result for the classical-quantum channel coding theorem \cite{Holevo-bounds,Holevo-bounds2}.
However, this research direction entered a period of stagnation in the 1980s.
In the 1990s, 
quantum information theory entered a new phase and 
was studied from a new viewpoint. 
For example, Schumacher introduced the concept of a typical sequence in a quantum system \cite{Schumacher}.
This idea brought us new developments and
enabled us to extend data compression to the quantum setting \cite{Schumacher}.
Based on this idea,
Holevo \cite{HoCh} and Schumacher and Westmoreland \cite{SW}
independently proved the classical-quantum channel coding theorem, which had been unsolved until that time. 

Unfortunately, a quantum operation in the framework of quantum information theory
is not necessarily available with the current technology.
Hence, these achievements remain more theoretical 
than classical channel coding theorem.
However, such theoretical results have, in a sense, brought us more practical results, as we shall see later.

Now, we give a formal statement of the quantum channel coding theorem for the classical-quantum channel $x \mapsto \rho_x$.
For this purpose, we introduce the von Neumann entropy 
$H(\rho):= - \Tr \rho \log \rho$ for a given density matrix $\rho$.
It is known that the von Neumann entropy is also concave
just as in the classical case.
When we employ the same classical-quantum channel $n$ times, 
the total classical-quantum channel 
is given as a map 
$x^n(\in {\cal X}^n) \mapsto 
\rho^{(n)}_{x^n}:=
\rho_{x_1}\otimes \cdots \otimes \rho_{x_n}$.
While an encoder is given as the same way as the classical case,
a decoder is defined in a different way because it is given by using a quantum measurement on the output quantum system ${\cal H}$.
The most general description of a quantum measurement on the output quantum system ${\cal H}$ is given by using a positive operator-valued measure
$D_n=\{\Pi_m\}_{m=1}^{M_n}$, in which, each $\Pi_m$ is a positive-semi definite matrix on ${\cal H}$ and the condition $\sum_{m=1}^{M_n} \Pi_m=I$ holds.
As explained in \cite[(4.7)]{Hay-book}\cite[(8.48)]{book2},
the decoding error probability is given as
$\epsilon(E_n,D_n):=
\frac{1}{M_n}\sum_{m=1}^{M_n} (1- \Tr \Pi_m \rho^{(n)}_{E_n(m)})$.
So, we can define 
the maximum transmission size
$M(\epsilon| \rho^{(n)}_{\cdot}):= \max_{(E_n,D_n)}\{|(E_n,D_n)| | \epsilon(E_n,D_n)\le \epsilon\} $.
On the other hand, the mutual information 
is defined as
$I(P_X,\rho_{\cdot}):=H(\sum_{x}P_X(x) \rho_x)-\sum_{x}P_X(x) H( \rho_x)$.
So, the maximum transmission
rate is characterized by the quantum channel coding theorem as follows
\begin{align}
\lim_{n \to \infty}
\frac{1}{n}\log M(\epsilon|\rho^{(n)}_{\cdot}) 
=\max_{P_X} I(P_X,\rho_{\cdot}), 
\quad 0<\epsilon <1.
\end{align}

To characterize the mutual information $I(P_X,\rho_{\cdot})$,
we denote the classical system ${\cal X}$
by using the quantum system ${\cal H}_X$ spanned by $|x\rangle$
and introduce the density matrix 
$\rho_{XY}:=\sum_{x \in {\cal X}}P_X(x) |x\rangle \langle x| \otimes \rho_x$ on the joint system ${\cal H}_X \otimes {\cal H}$
and 
the density matrix 
$\rho_{Y}:=\sum_{x \in {\cal X}}P_X(x) \rho_x$ on the quantum system ${\cal H}$.
In this notation, we regard $P_X$ as the density matrix 
$\sum_{x \in {\cal X}}P_X(x) |x\rangle \langle x| $ on ${\cal H}_X$.
Using the quantum relative entropy 
$D(\rho\|\sigma):= \Tr \rho (\log \rho-\log \sigma)$
between two density matrices $\rho$ and $\sigma$,
the mutual information is written as
\begin{align}
I(P_X,\rho_{\cdot})= 
D(P_X \otimes \rho_Y \| \rho_{XY})
=\min_{\sigma_Y}
D(P_X \otimes \sigma_Y \| \rho_{XY}).
\end{align}
So, the capacity is given by
\begin{align}
\max_{P_X}I(P_X,\rho_{\cdot}):= 
\max_{P_X}D(P_X \otimes \rho_Y \| \rho_{XY})
=\max_{P_X}\min_{\sigma_Y}
D(P_X \otimes \sigma_Y \| \rho_{XY}).
\end{align}

Here, it is necessary to discuss the relation between classical and quantum information theory.
For this purpose, we focus on information transmission via communication on an optical fiber.
When we employ coding in classical information theory, 
we choose a code based on classical information devices, which are
the input and the output of the classical channel shown in Fig. \ref{F2}.
In contrast,
when we employ coding in quantum information theory, 
we choose a code based on quantum information devices, which 
are the input and the output of the quantum channel shown in Fig. \ref{F3}.
In the case of Fig. \ref{F4}, we address the classical-quantum channel so 
that we focus on the output system as a quantum information device.
That is, the choice between classical and quantum information theory
is determined by the choice of a classical or quantum information device, 
respectively.

\section{Information Spectrum}\Label{S3}
The early stage of the development of finite block-length studies 
started from a completely different motivation and used
the information spectrum method introduced by Han and Verd\'{u}\cite{resolvability,Han1}.
Conventional studies in information theory
usually impose the iid or memoryless condition on the information source or the channel.
However, neither the information source nor the channel is usually independent in the actual case and 
they often have correlations.
Hence, information theory needed to be adapted for such a situation.

To resolve such a problem, Verd\'{u} and Han have discussed 
optimal performance in the context of several topics in classical information theory, including channel coding, by using the behavior of the logarithmic likelihood, as shown in Fig. \ref{F8}\cite{Verdu-Han}. 
However, they have discussed only the case when the block-length $n$ approaches infinity, and have not studied the case with finite block-length.
It is notable that this study clarified 
that the analysis of the iid case
can be reduced to the law of large numbers.
In this way, the information spectrum method has clarified the mathematical structures of many topics in information theory, which has worked as a silent trigger for further developments.

\begin{figure}[htbp]
\begin{center}
\scalebox{0.5}{\includegraphics[scale=1]{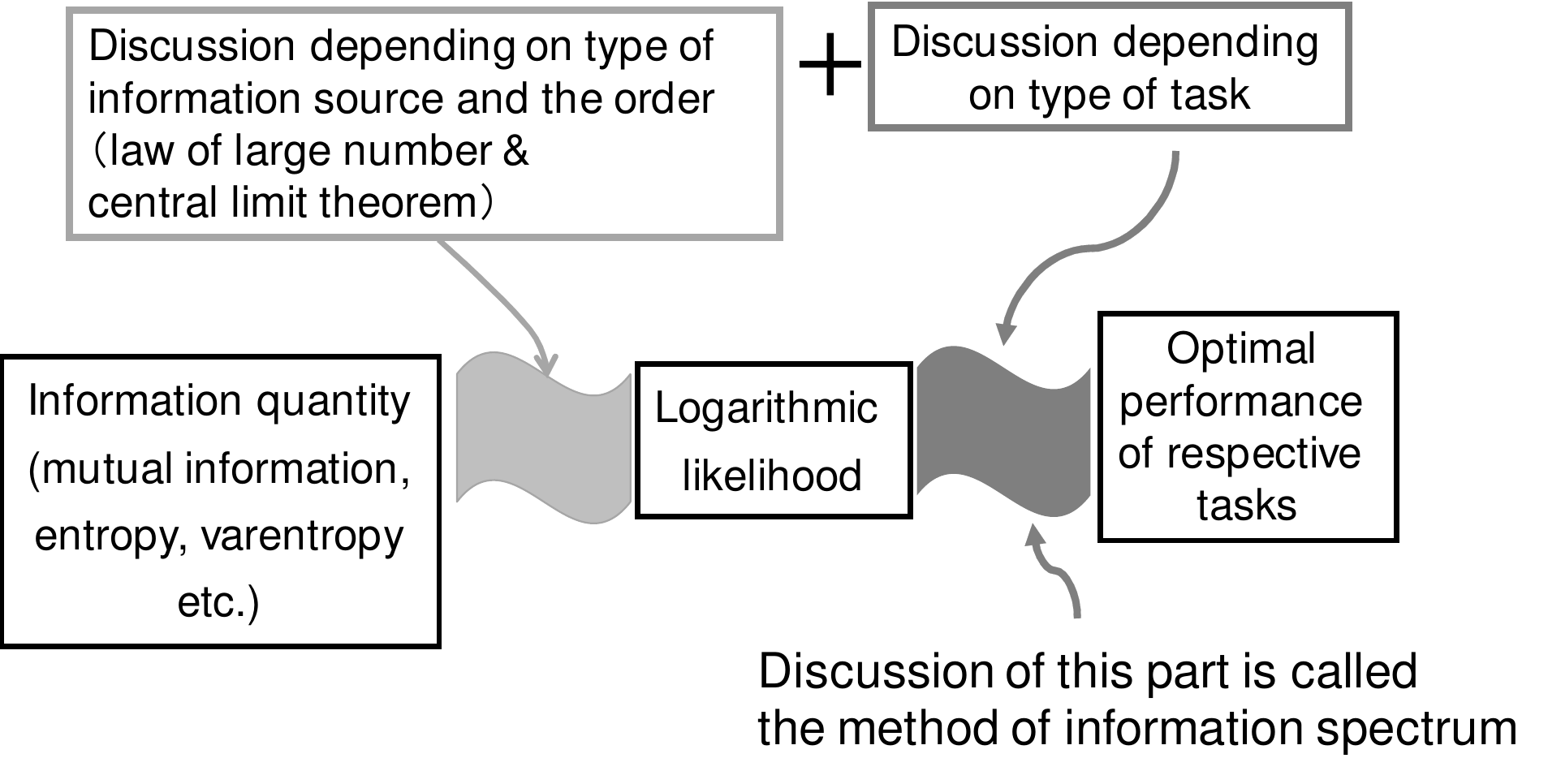}}
\end{center}
\caption{Structure of information spectrum:
The information spectrum method discusses the problem in steps.
One is the step to connect the information source and the behavior of the logarithmic likelihood.
The other is the step to connect the behavior of the logarithmic likelihood and the optimal performances in the respective tasks.}
\Label{F8}
\end{figure}%

Another important contribution of the information spectrum method
the connection of simple statistical hypothesis testing to many topics in classical information theory \cite{Han1}. 
Here, simple statistical hypothesis testing is the problem 
of deciding which candidate is the true distribution with an asymmetric treatment of two kinds of errors when two candidates for the true distribution are given.
In particular, the information spectrum method has revealed that 
the performances of 
data compression and uniform random number generation
are given by the behavior of the logarithmic likelihood. 

Here, we briefly discuss the idea of the information spectrum approach in the case of 
uniform random number generation.
Let ${\cal X}_n$ be the original system, where $n$ is an index.
The product set ${\cal X}^n$ is a typical example of this notation.
In uniform random number generation, we prepare another set ${\cal Y}_n$,
in which, we generate an approximate uniform random number $Y_n$.
In this formulation, we focus on the initial distribution $P_{X_n}$ on ${\cal X}_n$.
Then, our operation is given as a map $\phi_n$ from ${\cal X}_n$ to ${\cal Y}_n$.
The resultant distribution on ${\cal Y}_n$
is given as $P_{X_n}\circ \phi_n^{-1}(y):= \sum_{x : \phi_n(x)=y}P_{X_n}(x)$.
To discuss the quality of the resultant uniform random number,
we employ the uniform distribution 
$P_{{\cal Y}_n,\mix}(y):= \frac{1}{|{\cal Y}_n|}$ on ${\cal Y}_n$.
So, the error of the operation $\phi_n$ is given as
$\gamma(\phi_n):=
\frac{1}{2}\sum_{y \in {\cal Y}_n}
|P_{X_n}\circ \phi_n^{-1}(y)-P_{{\cal Y}_n,\mix}(y)|$.
Now, we define the maximum size of the uniform random number with error $\epsilon$ as
$S_n(\epsilon| P_{X_n}):= \max_{\phi_n} \{ |{\cal Y}_n|
| \gamma(\phi_n) \le \epsilon\}$.
Vembu and Verd\'{u} \cite[Section V]{Vembu-Verdu} showed that 
\begin{align}
\lim_{n \to \infty}
\frac{1}{n} \log S_n(\epsilon| P_{X_n})
=\sup_R \Big\{ 
R
\Big| \lim_{n \to \infty}
P_{X_n} \Big\{x \in {\cal X}_n \Big| -\frac{1}{n}\log P_{X_n}(x) \le R\Big\} 
\le \epsilon \Big\}.
\end{align}
This fact shows that the generation rate 
$\frac{1}{n} \log S_n(\epsilon| P_{X_n})$
is essentially described by the random variable $-\frac{1}{n}\log P_{X_n}(x)$.
When ${\cal X}_n$ is ${\cal X}^n$ and $P_{X_n}$ is the iid distribution $P_X^n$ of $P_X$,
the random variable $-\frac{1}{n}\log P_{X_n}(x)$ converges to 
the entropy $H(P_X)$ in probability due to the law of large numbers.
In the iid case, the generation rate equals the entropy $H(P_X)$.

In the channel coding case, we focus on a general conditional distribution 
$P_{Y_n|X_n}(y|x)$ as the channel.
Then, Verd\'{u} and Han \cite{Verdu-Han} derived the maximum transmission rate as
\begin{align}
\lim_{n \to \infty}
\frac{1}{n} \log M(\epsilon| P_{Y_n|X_n})
=
\sup_{\{P_{X_n}\}}
\sup_R \Big\{ 
R\Big| 
\lim_{n \to \infty}
P_{X_n,Y_n} \Big\{(x,y) \in {\cal X}_n\times {\cal Y}_n
\Big| \frac{1}{n}\log \frac{P_{Y_n|X_n}(y|x)}{P_{Y_n}(y)} \le R\Big\} 
\le \epsilon \Big\}.\Label{8-31-a}
\end{align}
Although we can derive the formula \eqref{9-1-11} from this general formulation,
it is not so easy because 
the above formula contains the maximization $\sup_{P_{X_n}}$ of the input distribution on the large system ${\cal X}_n$.
When the channel $P_{Y_n|X_n}$ is given as the additive channel with the additive noise distribution $P_{Z_n}$ as $P_{Y_n|X_n}(y|x)=P_{Z_n}(y-x)$,
the above formula can be simplified as 
\begin{align}
\lim_{n \to \infty}
\frac{1}{n} \log M(\epsilon| P_{Y_n|X_n})
=
\sup_R \Big\{ 
R\Big| \lim_{n \to \infty}
P_{Z_n} \Big\{z \in {\cal Z}_n
\Big| \frac{1}{n}(\log P_{Z_n}(z) + \log |{\cal Z}_n|)\le R\Big\} 
\le \epsilon \Big\}.\Label{8-31-ab}
\end{align}
Note that ${\cal Z}_n$ is the same set as ${\cal X}_n$ and ${\cal Y}_n$
when the channel is additive.

As already mentioned, the information spectrum approach was started as a result of 
a motivation different from the above.
When Han and Verd\'{u} \cite{resolvability} introduced this method, they considered identification codes, which were initially introduced by Ahlswede and Dueck \cite{AD}.
To resolve this problem, Han and Verd\'{u} introduced another problem---channel resolvability---
which discusses the approximation of a given output distribution by the input uniform distribution on a small subset.
That is, they consider 
\begin{align}
T(\epsilon|P_{Y_n|X_n})&:=
\max_{P_{X_n}}
T(\epsilon|P_{X_n},P_{Y_n|X_n}),
\end{align}
and
\begin{align}
&T(\epsilon|P_{X_n},P_{Y_n|X_n})\nonumber \\
:=&
\min_{{\cal T}_n} \min_{\phi_n}
\Bigg\{|{\cal T}_n| ~\Bigg| \frac{1}{2}
\sum_{y \in {\cal Y}_n} 
\bigg| 
\sum_{x\in {\cal X}_n} P_{Y_n|X_n}(y|x)P_{X_n}(x) - 
\sum_{x\in {\cal X}_n} P_{Y_n|X_n}(y|x)
\sum_{u:\phi_n(u)=x}
P_{{\cal T}_n,\mix}(x) 
\bigg| 
\le \epsilon
\Bigg\},
\end{align}
where
$\phi_n$ is chosen as a function from ${\cal T}_n$ to ${\cal X}_n$.
They showed that
\begin{align}
&\lim_{\epsilon \to 0}
\lim_{n \to \infty}
\frac{1}{n} \log T(\epsilon|P_{Y_n|X_n}) \nonumber \\
=&
\lim_{\epsilon \to 0}
\sup_{ \{P_{X_n}\}}
\sup_R \Big\{ 
R\Big| 
\lim_{n \to \infty}
P_{X_n,Y_n} \Big\{(x,y) \in {\cal X}_n\times {\cal Y}_n
\Big| \frac{1}{n}\log \frac{P_{Y_n|X_n}(y|x)}{P_{Y_n}(y)} \le R\Big\} 
\le \epsilon \Big\}.
\Label{9-3-1}
\end{align}
By considering this problem, they introduced the new concept of 
channel resolvability, which 
later played an important role in a completely different topic.

In the next stage, Nagaoka and the author extended the information spectrum method to the quantum case \cite{Nag-Hay,Hay-Nag}. 
In this extension, their contribution is not only the non-commutative extension but also the redevelopment of information theory.
In particular, they have given a deeper clarification of the explicit relation between
simple statistical hypothesis testing and channel coding, which is called the 
dependence test bound in the later study \cite[Remark 15]{Hay-Nag}.
In this context, Nagaoka \cite{Naga-EQIS} has developed another explicit relation between simple statistical hypothesis testing and channel coding, which is called the meta converse inequality\footnote{Unfortunately,
due to page limitations, the present paper cannot give a detailed derivation.
However, a detailed discussion is available in Section 4.6 of the book \cite{Hay-book}.}.
These two clarifications of the relation between 
simple statistical hypothesis testing and channel coding 
work as a preparation for the next step of finite-length analysis.

Now, to grasp the essence of these contributions, 
we revisit the classical setting
because the quantum situation is more complicated.
To explain the notation of classical hypothesis testing,
we consider testing between two distributions
$P_1$ and $P_0$ on the same system ${\cal X}$.
Generally, our testing method is written by using
a function $T$ from ${\cal X}$ to $[0,1]$ as follows.
When we observe $x \in {\cal X}$, 
we support $P_1$ with the probability $T(x)$, and support $P_0$ with the probability $1-T(x)$.
When the function $T$ takes values only in $\{0,1\}$,
our decision is deterministic.
In this problem, we have two types of error probability.
The first one is the probability for erroneously supporting $P_1$ while the true distribution is $P_0$, which is given as
$\alpha(T|P_0\|P_1):=\sum_{x\in {\cal X}}T(x) P_0(x)$.
The second one is the probability for erroneously supporting $P_0$ while the true distribution is $P_1$, which is given as
$\beta(T|P_0\|P_1):=\sum_{x\in {\cal X}}(1-T(x)) P_1(x)$.
Then, we consider the minimum second error probability under the constraint of a constant probability for the first error 
as
$\beta(\epsilon|P_0\|P_1):= \min_{T}\{
\beta(T|P_0\|P_1)| \alpha(T|P_0\|P_1) \le \epsilon\}$.

To overcome the problem with respect to $\sup_{P_{X_n}}$ in \eqref{8-31-a}, 
for a given channel $P_{Y|X}$,
Nagaoka \cite{Naga-EQIS} derived the meta converse inequality:
\begin{align}
M(\epsilon| P_{Y|X}) \le 
\max_{P_X}
\beta(\epsilon|P_{XY}\|P_{X}\times Q_Y)^{-1}\Label{8-31-6}
\end{align}
for any distribution $Q_Y$ on ${\cal Y}$.

Also, the author and Nagaoka derived the dependence test bound as follows \cite[Remark 15]{Hay-Nag}.
For a given distribution on $P_X$ on ${\cal X}$
and a positive integer $N$,
there exists a code $(E,D)$ such that $|(E,D)|=N$ \footnote{In the quantum case, they found a slightly weaker inequality.
However, we can trivially derive \eqref{8-31-5} from their derivation in the commutative case.}
\begin{align}
\epsilon(E,D) \le
\epsilon+ N \beta(\epsilon|P_{XY}\|P_{X}\times P_Y).\Label{8-31-5B}
\end{align}
That is, for any $\delta>0$ and $\epsilon>0$, we have
\begin{align}
M(\epsilon+\delta | P_{Y|X}) \ge
\max_{P_X} 
\delta \beta(\epsilon|P_{XY}\|P_{X}\times Q_Y)^{-1}.\Label{8-31-5}
\end{align}
Here, \eqref{8-31-5} follows from \eqref{8-31-5B} by putting $\delta= N \beta(\epsilon|P_{XY}\|P_{X}\times P_Y)$. 

Then, using \eqref{8-31-5}, 
the author and Nagaoka derived the $\ge$ part of \eqref{8-31-a} 
including the quantum extension.
Also,
using \eqref{8-31-6},
the author and Nagaoka derived another expression for 
\eqref{8-31-a}:
\begin{align}
&\lim_{n \to \infty}
\frac{1}{n} \log M(\epsilon| P_{Y_n|X_n}) \nonumber \\
=&
\inf_{\{Q_{Y_n}\}}
\sup_{\{P_{X_n}\}}
\sup_R \bigg\{ 
R\bigg| 
\lim_{n \to \infty}
P_{X_n,Y_n} \bigg\{(x,y) \in {\cal X}_n\times {\cal Y}_n
\bigg| \frac{1}{n}\log \frac{P_{Y_n|X_n}(y|x)}{Q_{Y_n}(y)} \le R\bigg\} 
\le \epsilon \bigg\}.\Label{8-31-b}
\end{align}
While \eqref{8-31-b} seems more complicated than \eqref{8-31-a},
\eqref{8-31-b} is more useful for proving the impossibility part for the following reason.
In \eqref{8-31-a}, the distribution $P_{Y_n}$ has a complicated form in general.
Hence, it is quite difficult to evaluate the behavior of 
$\frac{1}{n}\log \frac{P_{Y_n|X_n}(y|x)}{P_{Y_n}(y)} $.
When we derive the upper bound of 
$\lim_{n \to \infty}\frac{1}{n} \log M(\epsilon| P_{Y_n|X_n})$,
it is enough to consider the case with a special $Q_{Y_n}$.
That is, $Q_{Y_n}$ can be chosen to be a distribution for iid random variables so that
the random variable $\frac{1}{n}\log \frac{P_{Y_n|X_n=x}(y)}{Q_{Y_n}(y)} $ can be factorized.
Then, the impossibility part of the channel coding theorem
can be easily shown via \eqref{8-31-b}.

Indeed, since the classical case is not so complicated,
it is possible to recover several important results from \eqref{8-31-a}.
However, use of the formula \eqref{8-31-b} is needed in the quantum case
because everything becomes more complicated.

\section{Folklore in Source Coding}\Label{S4}
When the information source is subject to the iid distribution $P_X^n$ of $P_X$,
the compression rate and the uniform random number generation rate have the same value of $H(P_X)$ asymptotically.
Hence, we can expect that the data compressed up to the entropy rate 
$H(P_X)$ would be the uniform random number.
However, this argument does not work as a proof of the statement, so
this conjecture has the status of folklore in source coding,
and its validity remained unconfirmed for a long time.

Han \cite{Han:Folk} tackled this problem by using the method of information spectrum.
Han focused on the normalized relative entropy 
$\frac{1}{n}D(P_{X}^n \circ \phi_n^{-1}\| P_{{\cal Z}_n,\mix})$
as the criterion to measure 
the difference of the generated random number from a uniform random number,
and showed that the folklore  in source coding is valid \cite{Han:Folk}.
However, the normalized relative entropy is too loose a criterion to guarantee 
the quality of the uniform random number
because it is possible to distinguish a generated random number from a truly uniform random number even though the random number is considered to be uniform by this criterion.
In particular, when a random number is used for cryptography,
we need to employ a more rigorous criterion to judge the quality of its uniformity.

In contrast, the criterion $\gamma(\phi_n) $ is the most popular criterion
which gives the statistical distinguishability between a truly uniform random number and a given random number \cite{R-K}.
That is, when this criterion takes the value $0$,
the random number must be truly uniform.
Hence, when we use a random number for cryptography, 
we need to accept only a random number passing this criterion.
Also, Han \cite{Han:Folk} has proved that the folklore conjecture in source coding is not valid
when we adopt the variational distance as our criterion. 

On the other hand, to clarify the incompatibility between data compression and uniform random number generation,
the author \cite{Hay2} developed a theory for finite-block-length codes for both topics.
In this analysis, 
he applied the method of information spectrum  to the second-order $\sqrt{n}$ term, as shown in Fig. \ref{F8}.
That is, 
by using the varentropy $V(P_X):=
\sum_{x\in {\cal X}} P_X(x) (- \log P_X(x)-H(P_X))^2$,
the central limit theorem guarantees that 
\begin{align}
P_X^n \{x^n \in {\cal X}^n| 
(\log P_{X}^n(x^n)- nH(P_X))/\sqrt{n} \le \epsilon\}
= \sqrt{V(P_X)}\Phi^{-1}(\epsilon),
\Label{12-9-1}
\end{align}
where the cumulative distribution function $\Phi$ of the standard Gaussian distribution is defined as
$\Phi(a):= \frac{1}{\sqrt{2\pi}}\int_{-\infty}^a e^{-\frac{t^2}{2}}dt$.
So, the generation length $\log S(\epsilon |P_X^n)$ is asymptotically expanded as
\begin{align}
\log S(\epsilon |P_X^n)
= n H(P_X)  + \sqrt{n}\sqrt{V(P_X)}\Phi^{-1}(\epsilon)
+ o(\sqrt{n}).\Label{8-31-9}
\end{align}

Now, we consider data compression, in which we define
the  minimum compressed size 
$ R(\epsilon |P_X^n)$ with decoding error $\epsilon$
in the same way.
Then, the asymptotic expansion is \cite{strassen,Hay2}
\begin{align}
\log R(\epsilon |P_X^n)= n H(P_X)  - \sqrt{n} \sqrt{V(P_X)}\Phi^{-1}(\epsilon)
+ o(\sqrt{n}).\Label{8-31-10}
\end{align}
That is, whenthe converted length has the asymptotic expansion
$ n H(P_X) - \sqrt{n}\sqrt{V(P_X)} R $,
the errors of both settings are illustrated in Fig. \ref{F7}. 

\begin{figure}[htbp]
\begin{center}
\scalebox{0.5}{\includegraphics[scale=1]{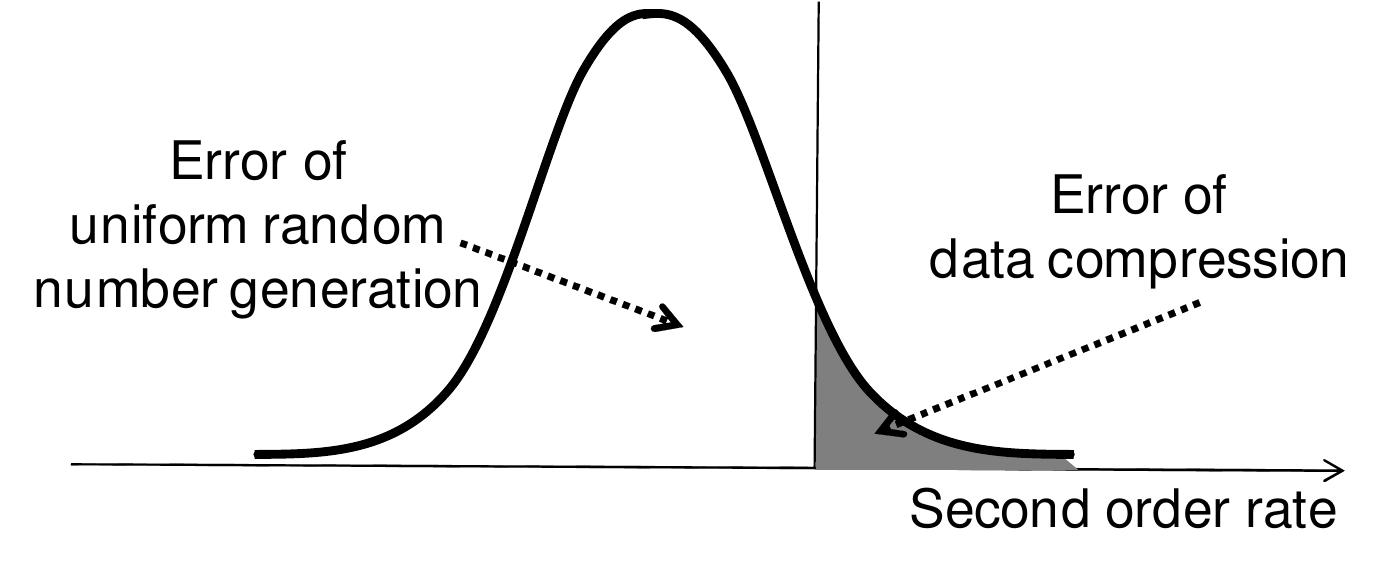}}
\end{center}
\caption{Asymptotic trade-off relation between errors of data compression and 
uniform random number generation:
When we focus on the second-order coding rate,
the minimum error of data compression is the probability of the exclusive event
of the minimum error of uniform random number generation.}
\Label{F7}
\end{figure}%

Now, we fix the conversion rate up to the second-order $\frac{1}{\sqrt{n}}$.
When we apply an operation from the system ${\cal X}^n$ 
to a system with size $e^{nH(P_X)+\sqrt{n}R}$,
the sum of the errors of 
the data compression and the uniform random number generation almost equals to $1$.
This trade-off relation shows that 
data compression and uniform random number generation 
are incompatible to each other.
Indeed, since the task of data compression has the direction opposite to that of uniform random number generation,
the second-order analysis explicitly clarifies that there is a trade-off relation for their errors rather than compatibility.

Although the evaluation of optimal performance up to the second-order coefficient gives 
an approximation of the finite-length analysis,
it also shows the existence of their  trade-off relation.
This application shows the importance of the second-order analysis.
Because the evaluation of the uniformity of a random number is closely related to
security, 
this type evaluation has been applied to security analysis \cite{Watanabe}.
This trade-off relation also plays an important role 
when we use the compressed data as the scramble random variable for another piece of information \cite{H-RM}.


\section{Quantum cryptography}\Label{S5}
\subsection{Single-photon pulse without noise}\Label{S5A}
Section \ref{S2} has explained that the problem of the ultimate performance of optical communication 
can be treated as quantum channel coding.
When the communication media has quantum properties,
it opens the possibility of a new communication style that cannot be realized with the preceding technology.
Quantum cryptography was proposed by Bennett and Brassard \cite{BB84} in 1984 as
a technology to distribute secure keys by using quantum media.
Even when the key is eavesdropped during the distribution,
this method enables us to detect the existence of the eavesdropper with high probability.
Hence, this method realizes secure key distribution,
and is called quantum key distribution (QKD).

Now, we explain the original QKD protocol based on single-photon transmission.
In the QKD, 
the sender, Alice, needs to generate four kinds of states in the two-dimensional system $\bC^2$, namely,
$|0\rangle,|1\rangle,$ and $|\pm \rangle:= 
\frac{1}{\sqrt{2}}(|0\rangle\pm|1\rangle)$\footnote{In the study of cryptography,
We call the authorized sender, the authorized receiver, and the eavesdropper
Alice, Bob, and Eve, respectively.}.
Here, $\{|0\rangle,|1\rangle\}$ is called the bit basis,
and $\{|\pm\rangle\}$ is called the phase basis.
Also, the receiver, Bob, needs to measure the received quantum state by using either the bit basis or the phase basis.

The original QKD protocol \cite{BB84} is the following.
\begin{description}
\item[(1)] [Preparation] Alice randomly chooses one of four states, and sends it to Bob.
\item[(2)] [Transmission] Bob randomly chooses one of two bases, and measures the received state using the chosen basis.
Alice and Bob repeat Steps (1) and (2) several times.
\item[(3)] [Detection] Alice and Bob exchange their basis information via a public 
channel, and they discard bits with disagreed  bases.
\item[(4)] [Error check] Alice and Bob randomly choose check bits from among the remaining bits,
and they exchange their values via a public channel.
If they find an error, 
they stop the protocol because the error might be caused by eavesdropping.
Otherwise, they use the remaining bits as keys, which are called {\it raw} keys.
\end{description}

In this protocol, if the eavesdropper, Eve, performs a measurement during transmission,
the quantum state would be destroyed with non-negligible probability
because she does not know the basis of the transmitted quantum state a priori.
When the number of qubits measured by Eve is not so small,
Alice and Bob will find disagreements in step (4).
So, the existence of eavesdropping will be discovered by Alice and Bob with high probability.

\subsection{Random hash functions}\Label{S5A2}
The original protocol supposes noiseless quantum communication by a single photon.
So, the raw keys are not necessarily secure when the channel has noise.
To realize secure communication even with a noisy channel, 
we need a method to generate secure keys from keys partially leaked to Eve.
Such a process is called privacy amplification.
In this process, we apply a hash function, which maps from a larger set to a smaller set.
In the security analysis, 
we often employ a hash function
whose choice is determined by a random variable (a random hash function).
A typical class of random hash functions is the following class.
A random hash function $f_R$ 
from $\bF_2^{n}$ to $\bF_2^{m}$
is called universal$_2$ \cite{Carter,WC81} when 
\begin{align}
\Pr \{f_R (x)= f_R (x')\} \le 2^{-m}
\end{align}
for distinct elements $x$ and $x'$ in $\bF_2^{n}$.
A typical example of a surjective universal$_2$ hash function is 
the concatenated Toeplitz matrix, which is given as follows.
When an $m\times (n-m)$ matrix $T_R=(T_{i,j}) $ is given as
$T_{i,j}=R_{i+j-1}$ by using $n-1$ random variables $R_j$,
it is called a Toeplitz matrix.
Let ${\cal T}=\{T_R\,|\, r\in I\}$ be the set of all $m\times (n-m)$ Toeplitz matrices.
Then let $M_r=(T_R,I_{m})$ be an $m\times n$ matrix defined by a concatenation of 
$T_R$ and the $m$-dimensional identity matrix $I_m$.
Then, the concatenated Toeplitz matrix $M_R$ maps 
an input $x\in \bF_2^n$ to the output $y=M_r x \in \bF_2^m$. 
The concatenated Toeplitz matrix $M_R$ is universal$_2$ when $R$ is a uniform random number. (For a proof, see, e.g., \cite[Appendix II]{Haya5}.)

This class can be relaxed as follows. 
A random hash function $f_R$ from $\bF_2^{n}$ to $\bF_2^{m}$ is called
$\delta$-almost universal$_2$ when 
\begin{align}
\Pr \{f_R (x)= f_R (x')\} \le \delta 2^{-m}
\end{align}
for distinct elements $x$ and $x'$ in $\bF_2^{n}$.
Here, $\Pr \{C \} $ expresses the probability that the condition $C$ holds.
When $\delta=1$, it is universal$_2$.
Here, $R$ denotes the random variable identifying the hash function.
When a random hash function $f_R$ is linear,
it is $\delta$-almost universal$_2$
if and only if
\begin{align}
\Pr \{ x \in \Ker f_R \} \le \delta 2^{-m}
\end{align}
for any non-zero element $x \in \bF_2^{n}$.
Here, $\Ker f$ is the kernel of the linear function $f$.
Considering the space $(\Ker f)^{\perp}$ orthogonal to $\Ker f$ in $\bF_2^{n}$, we introduce another class of random hash functions.
A linear random surjective hash function $f_R$ 
from $\bF_2^{n}$ to $\bF_2^{m}$
is called $\delta$-almost dual universal$_2$
when 
\begin{align}
\Pr \{ x \in (\Ker f_R)^{\perp} \} \le \delta 2^{-n+m}
\end{align}
for any non-zero element $x \in \bF_2^{n}$.
As examples of $\delta$-almost dual universal$_2$ hash functions,
the paper \cite{HT2} proposed hash functions whose calculation complexity 
and random seeds are smaller than existing functions for practical use, as shown in Fig. \ref{F6}.

When $R$ is not a uniform random number the above concatenated Toeplitz matrix $M_R$ is not universal$_2$; fortunately, it is $\delta$-almost dual universal$_2$. 
So, we can evaluate security in the framework of $\delta$-almost dual universal$_2$ hash functions.
That is, for a realistic setting,
the concept of $\delta$-almost dual universal$_2$ works well.
Note that there exists a $2$-almost universal$_2$ hash function whose resultant random number is insecure (Fig. \ref{F6}).
Hence, the concept of $\delta$-almost dual universal$_2$ is more useful than 
$\delta$-almost universal$_2$.

\begin{figure}[htbp]
\begin{center}
\scalebox{0.5}{\includegraphics[scale=1]{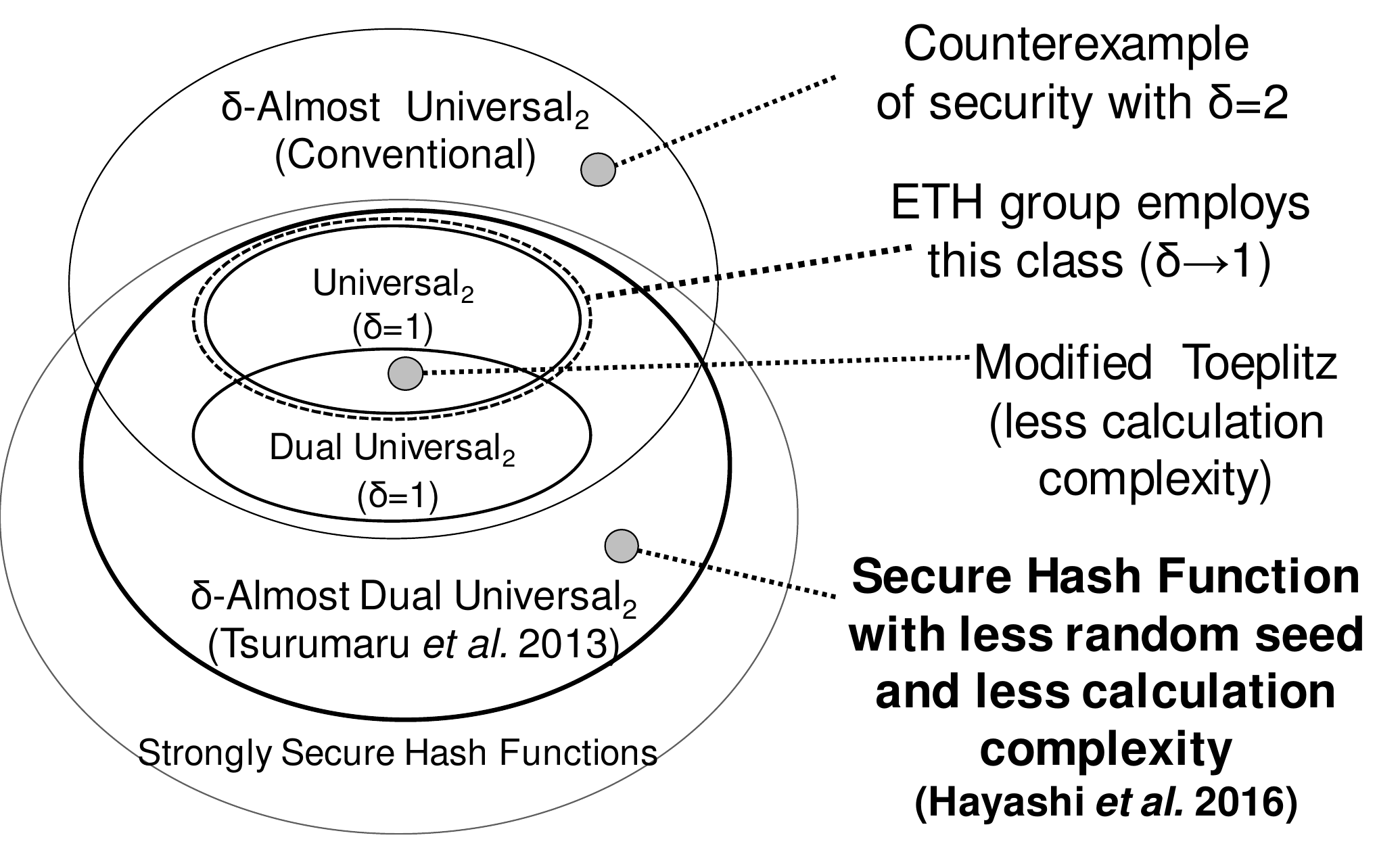}}
\end{center}
\caption{Classes of (dual) universal$_2$ hash functions and security:
A hash function is used to realize privacy amplification.
This picture shows the relations between classes of hash functions and security.
In cryptography theory, 
strong security is considered a requirement for a hash function \cite{R-K}.
The class of universal$_2$ hash functions was proposed in \cite{Carter,WC81}.
Using the leftover hash lemma \cite{BBCM,ILL},
Renner \cite{Renner1} proposed to use this class for quantum cryptography.
Tomamichel et al. \cite{TSSR11} 
proposed to use the class of $\delta$-almost universal$_2$ hash functions
when $\delta$ is close to $1$.
Tsurumaru et al. \cite{Tsuru} proposed the use 
of $\delta$-almost dual universal$_2$ hash functions
when $\delta$ is constant or increases polynomially.
As an example of a $\delta$-almost dual universal$_2$ hash function,
the author with his collaborators \cite{HT2} constructed a secure Hash Function with a less random seed
and less calculation.
Although the security analysis in \cite{TWGR} is based on universal$_2$ hash functions,
that in \cite{H-QKD2,H-N,HT1} is based on $\delta$-almost dual universal$_2$ hash functions.}
\Label{F6}
\end{figure}%

\subsection{Single-photon pulse with noise}\Label{S5B}
To realize the security even with a noisy quantum channel,
we need to modify the original QKD protocol.
Since this modified protocol is related to error correction, 
finite-length analysis plays an important role 
to guarantee the security of the real QKD system.
Here, for simplicity, we discuss only the finite-length security analysis 
with  the Gaussian approximation.

The modified QKD protocol is the following.
Steps (1), (2), and (3) are the same as in the original.
\begin{description}
\item[(4)] [Error estimation] Alice and Bob randomly choose check bits from among the remaining bits,
and they exchange their values via a public channel.


\item[(*)] In the following, we give a protocol for the bit basis.
Here, we denote the number of remaining bits with the bit basis measurement 
by $n$,
and
we denote the numbers of check bits with the phase and bit basis measurements by $l$ and $l'$.
We denote the numbers of observed errors among 
check bits in the phase and bit basis measurements by $c$ and $c'$.

\item[(5)] [Error correction]
Alice and Bob apply error correction based on a $k$-dimensional subspace $C$ and obtain $k$ corrected bits.
That is, Alice sends her syndrome to Bob via a public channel,
and Bob corrects his error.
Here, the length $k$ and a code $C$ are chosen by 
the observed error rate $\frac{c'}{l'}$ with the bit basis measurement.

\item[(6)] [Privacy amplification]
Alice and Bob apply a $\delta$-almost dual universal2 hash function from $\bF_2^k$ to $\bF_2^{k-\bar{k}}$.
This protocol sacrifices $\bar{k}$ bits, which is called the sacrifice bit length and is determined by the observed error rate $\frac{c}{l}$ with the phase basis measurement.
Then, Alice and Bob obtain final keys with length $s:=k-\bar{k}$.
\end{description}

To perform the finite-length security analysis approximately,
we consider the following items.
\begin{description}
\item[(i)]
The virtual decoding phase error probability of a code $C$ with an arbitrary decoder 
gives the amount of leaked information with privacy amplification by a hash function whose kernel is $C^{\perp}$.
In this correspondence,
the privacy amplification in the bit basis 
by a $\delta$-almost dual universal$_2$ hash function $\bF_2^k$ to $\bF_2^{k-\bar{k}}$
essentially realizes 
an error correction code in the phase basis
whose parity check matrix is a $\delta$-almost universal$_2$ hash function
from $\bF_2^n$ to $\bF_2^{\bar{k}}$\cite[Lemmas 2 \& 4]{H-QKD}\cite[Theorem 2]{H-QKD2}\cite[(54)]{Tsuru}\cite[Section 9.4.3]{book2}\cite[Section 5.6.2]{book3}\footnote{To explain this point, we need to discuss a $\delta$-almost universal$_2$ hash function for $\bF_2^n/C^{\perp}$, which requires more work.
To avoid this difficulty, we give only a simplified discussion here.}.

\item[(ii)] 
When the total number of bits is $n+l$, the total number of errors is $b$,
and we randomly choose $l$ bits as the observed bits,
the number of observed errors $c$ is subject to the hypergeometric distribution
$P_{b}(c):=\frac{{l \choose c} {n \choose b-c}}{{n+l \choose b}}$.
So, the value $(c-\frac{lb}{n+l})/\sqrt{l} $ 
approximately obeys the Gaussian distribution with variance 
$ \frac{bn(n+l-b)}{(n+l)^2(n+l-1)}$.

\item[(iii)] 
When the parity check matrix is given by a $\delta$-almost universal$_2$ hash function from $\bF_2^{n}$ to $\bF_2^{\bar{k}}$,
the decoder is the minimum distance decoder,
and
the support of the distribution $P_{Z^n}$ of errors on $\bF_2^n$
is included in the set $\{x^n \in \bF_2^n| |x^n|=b-c \}$,
the average decoding error probability is evaluated as 
\begin{align}
\rE_R \epsilon(E_{f_R},D_{f_R,\min}|P_{Z^n})
\le \delta e^{n h((b-c)/n)-\bar{k}},
\end{align}
where
$\rE_R$ denotes the expectation with respect to the random variable $R$\cite[Lemma 1]{H-QKD}\cite[Theorem 3]{H-QKD2}\cite[(37)]{Tsuru}.

\item[(iv)] 
The real distribution of error in the phase basis
for $n$ remaining qubits with the bit basis measurement 
and $l$ check qubits with the phase basis measurement
($n+l$ qubits in total)
is written as a probabilistic mixture of 
distributions $P_{\bar{k}}$, where $P_{\bar{k}}$ is a distribution on $\{x^n \in \bF_2^{n+l}| |x^n|=\bar{k} \}$\cite[Section IV-B]{H-QKD}\cite[Section III-C]{H-QKD2}\cite[(18)]{HT1}.
(Any distribution on $\bF_2^n$ satisfies this condition.
In the memoryless case, the coefficients form a binomial distribution.)
\end{description}

To give our security criterion,
we denote the information transmitted via the public channel by $u$,
and introduce its distribution $P_{{\rm pub}}$. 
Depending on the public information $u$,
we denote the state on the composite system of Alice's and Eve's systems,
the state on Alice's system,
the state on Eve's system,
and the length of the final key length
by $\rho_{A,E|u}$, $\rho_{A| u}$, $\rho_{E|u}$, and $s(u)$, respectively.
We denote the completely mixed state with length $s(u)$
by $\rho_{A,\mix| s(u)}$.
Then, similar to 
the security criterion is given in \cite[(3)]{HT1}
\begin{align}
\frac{1}{2}
\sum_{u} P_{{\rm pub}}(u)
\| \rho_{A,E|u}- \rho_{A,\mix| s(u)}\otimes \rho_{E|u}\|_1.
\Label{9-1-1}
\end{align}
Now, as a security condition, we impose the condition that 
\eqref{9-1-1} is smaller than $\epsilon$.

Combining the above four items, 
depending on $c$,
we can derive the sacrifice bit length $\bar{k}(c)$. 
Although the exact formula of $\bar{k}(c)$ is complicated, it can be asymptotically expanded as \cite[(53)]{HT1} 
\begin{align}
\bar{k}(c)= nh(\frac{c}{l})
-  \frac{\sqrt{n}}{2}h'(\frac{c}{l})\sqrt{
\frac{c}{l}(1-\frac{c}{l})(1+\frac{l}{n})\frac{n}{l}}
\Phi^{-1}(\frac{\epsilon^2}{2})
+o(\sqrt{n}).
\end{align}

Here, we should remark that this security analysis does not assume the memoryless condition for the quantum channel.
To avoid this assumption, we introduce a random permutation and the effect of random sampling, which allows us to consider that the errors in both bases are subject to the hypergeometric distribution.
However, due to the required property of hash functions, we do not need to apply the random permutation in the real protocol. 
That is, we need to apply only random sampling to estimate the error rates of the phase basis.

Here, we need to consider the reliability, that is, the agreement of the final keys.
For this purpose, we need to attach a key verification step as follows \cite[Section VIII]{Fung}.

\begin{description}
\item[(7)] [Key verification] 
Alice and Bob apply a universal$_2$ hash function from $\bF_2^{k-\bar{k}}$ to $\bF_2^{\hat{k}}$ to the final keys.
They exchange their results via a public channel.
They discard their final $\hat{k}$ bits if they do not agree.
Otherwise, they consider that their remaining keys agree.
\end{description}

However, the amount of leaked information for the final keys 
cannot be estimated by a similar method.
So, the security analysis is more important than the agreement of the keys.

\subsection{Weak coherent pulse with noise}\Label{S5C}
Next, we discuss a weak coherent pulse with noise, whose device is illustrated in Fig. \ref{F6B}.
Since the above protocol assumes single-photon pulses,
when the pulse contains multiple photons even occasionally,
the above protocol cannot guarantee security.
Since it is quite difficult to generate a single-photon pulse,
we usually employ weak coherent pulses with phase randomization,
whose states are written as
$\sum_{n=0}^{\infty}e^{-\mu}\frac{\mu^n}{n\!} | n \rangle \langle n |$,
where $\mu$ is called the intensity.
That is, weak coherent pulses contain multiple-photon pulses, as shown in Fig. \ref{F9}.
In this case, 
there are several multiple-photon pulses among $n$ received pulses. 
In optical communication, 
only a small fraction of pulses arrive at the receiver side.
That is, the ratio of multiple-photon states
of Alice's side is different 
from that of Bob's side.
This is because the detection ratio on Bob's side depends on the number of photons.

As the first step in the security analysis,
we need to estimate the ratios of 
vacuum pulses, single-photon pulses, and multiple-photon pulses among 
$n$ received pulses.
Indeed, there is a possibility that Bob erroneously detects the pulse even with a vacuum pulse.
To obtain this estimate, we remark that the ratio of multiple-photon pulses depends on the intensity $\mu$.
Hence, it is possible to estimate 
the detection ratios of 
vacuum pulses, single-photon pulses, and multiple-photon at Bob's side
from the detection ratios of more than 3 different intensities,
which are obtained by solving simultaneous equations \cite{decoy1,decoy2,decoy3,Ma05,Wang05,H1,decoy4,ODI}.
Observing the error rate of each pulse depending on the intensity and the basis, we can estimate the error rates of both bases for 
vacuum pulses, single-photon pulses, and multiple-photons. 
This idea is called the decoy method.
Based on this discussion, we change steps (1), (2), (3), and (4).
However, we do not need to change steps (5) and (6), in which we choose the error correcting code and the sacrifice bit length.

As the second step of the security analysis,
when $n$ received pulses are composed of
$n_0$ vacuum pulses, $n_1$ single-photon pulses,
and $n_2$ multiple-photon pulses,
we need to estimate the leaked information after the privacy amplification
with sacrifice bit length $\bar{k}$.
In the current case, 
we replace items (i) and (iii) by the following.
\begin{description}
\item[(i')]
When $n$ received pulses are composed of
$n_0$ vacuum pulses, $n_1$ single-photon pulses,
and 
$n_2$ multiple-photon pulses,
then, $n_0$ vacuum pulses are converted to noiseless single-photon pulses
and 
$n_2$ multiple-photon pulses
are converted to noiseless single-photon pulses
whose error distribution is the uniform distribution \cite[Section III-B]{H-QKD2}.
Then, we have the same statement as (i). 

\item[(iii')] 
Assume that the parity check matrix is given by a $\delta$-almost universal$_2$ hash function from $\bF_2^{n}$ to $\bF_2^{\bar{k}}$.
We also make an assumption for the distribution $P_{Z^n}$
on $\bF_2^n=\bF_2^{n_0+n_1+n_2}$;
$n_0$ bits have no error,
there are $t_1$ errors among the $n_1$ bits, and
the distribution of errors on the $n_2$ bits
is the uniform distribution.
So, the decoder $\Gamma([y])$ is defined as
\begin{align}
\Gamma([y]):= \argmin_{x^n \in [y]:(*)}
\|x^n\|,
\end{align}
where $(*)$ is the condition that 
all of entries among the above $n_0$ bits are $0$,
and $\|x^n\|$ is the number of bits with entry $1$ among the above $n_1$ bits.
Then, the average decoding error probability is evaluated as \cite[Theorem 3]{H-QKD2}
\begin{align}
\rE_R \epsilon(E_{f_R},D_{f_R,\min}|P_{Z^n})
\le \delta e^{n_1 h(t_1/n_1)+ n_2-\bar{k}}.
\end{align}
\end{description}

Finally, we combine the original items (ii) and (iv) with the above modified items (i') and (iii'). 
However,
due to the complicated estimation process for the partition $n_0,n_1,n_2$ of 
$n$ qubits,
we need a very complicated discussion.
Based on such an analysis, after long calculation,
we obtain a formula for the sacrifice bit length, as shown in Fig. \ref{F5}.

\newpage

\begin{figure}[htbp]
    \begin{tabular}{cc}
      \begin{minipage}[t]{0.45\hsize}

        \centering
\scalebox{0.5}{\includegraphics[scale=0.3]{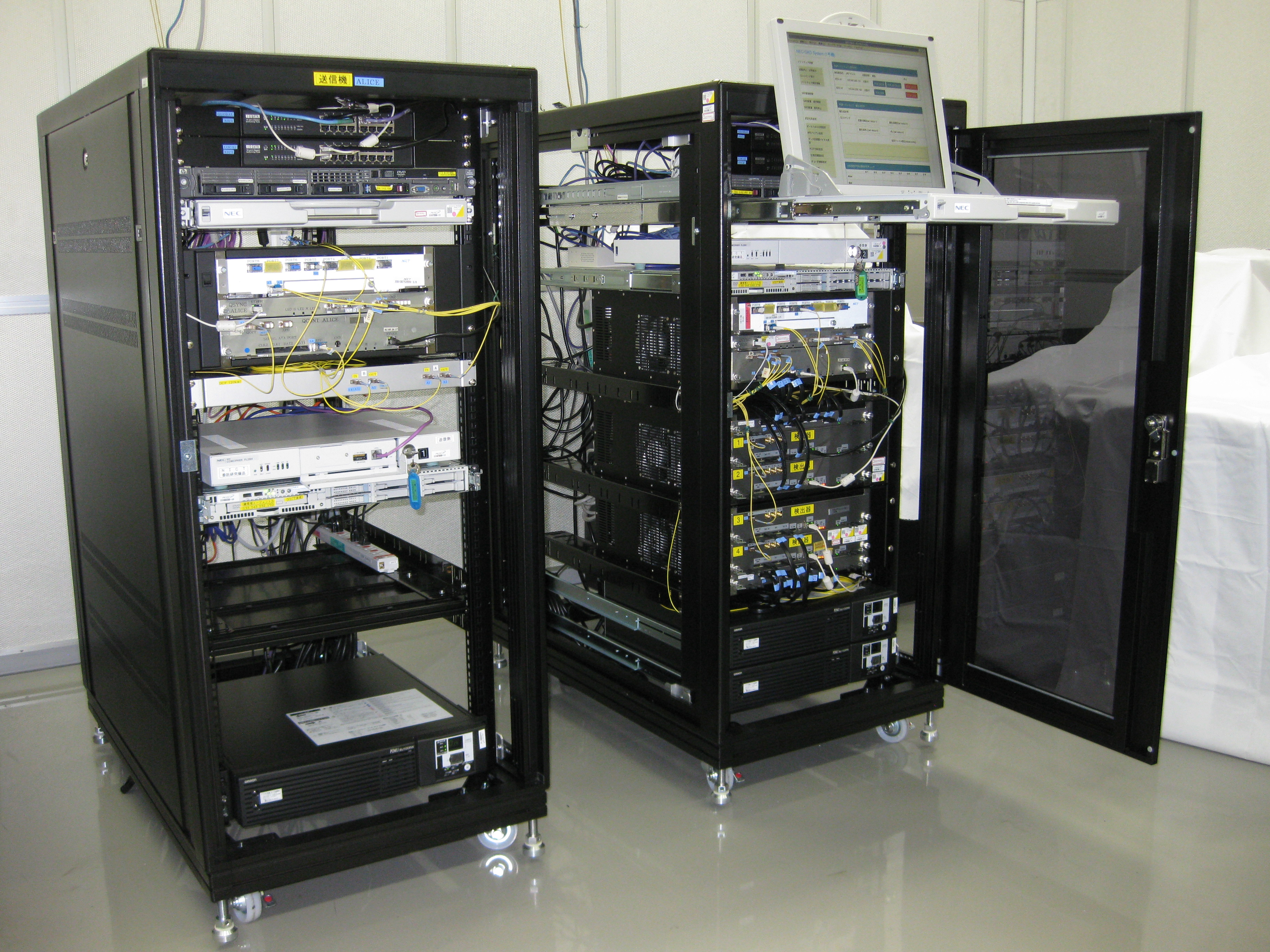}}
\caption{QKD system developed by NEC. Copyright (2015) by NEC:
This device was used for a long-term evaluation 
demonstration in 2015 by the ``Cyber Security Factory" (core facility for counter-cyber-attack activities in NEC) \cite{Nec}.
}
\Label{F6B}
      \end{minipage} &
      \begin{minipage}[t]{0.45\hsize}
        \centering
\scalebox{0.5}{\includegraphics[scale=1]{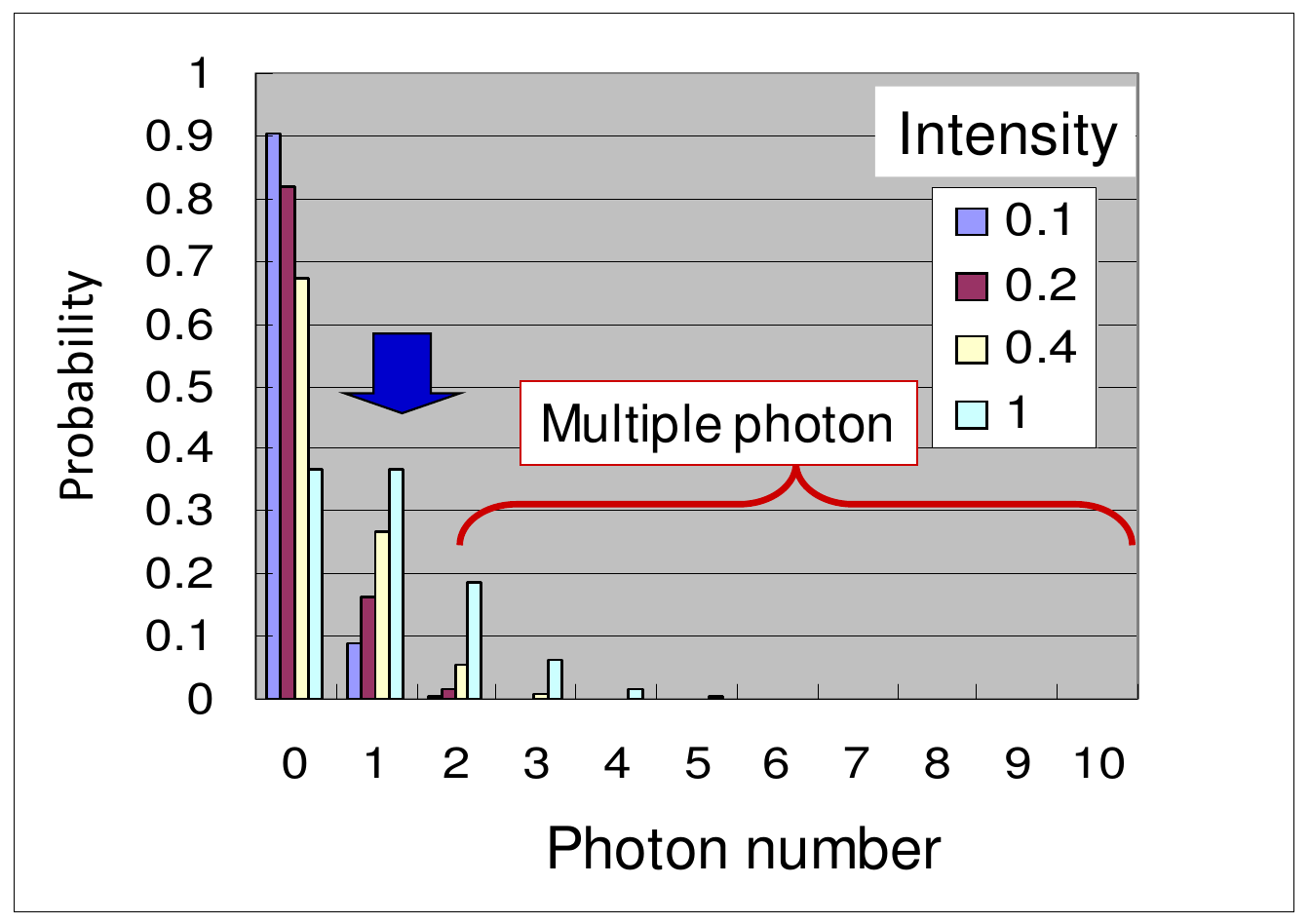}} 
\caption{Multiple photons in a weak coherent pulse:
A weak coherent pulse contains multiple photons with a certain probability, which depends on the intensity of the pulse.}
\Label{F9}
      \end{minipage}
    \end{tabular}
  \end{figure}

\begin{figure}[htbp]
\begin{tabular}{cc}
\begin{minipage}{0.7\hsize}
\begin{center}
\includegraphics[width=0.7\textwidth]{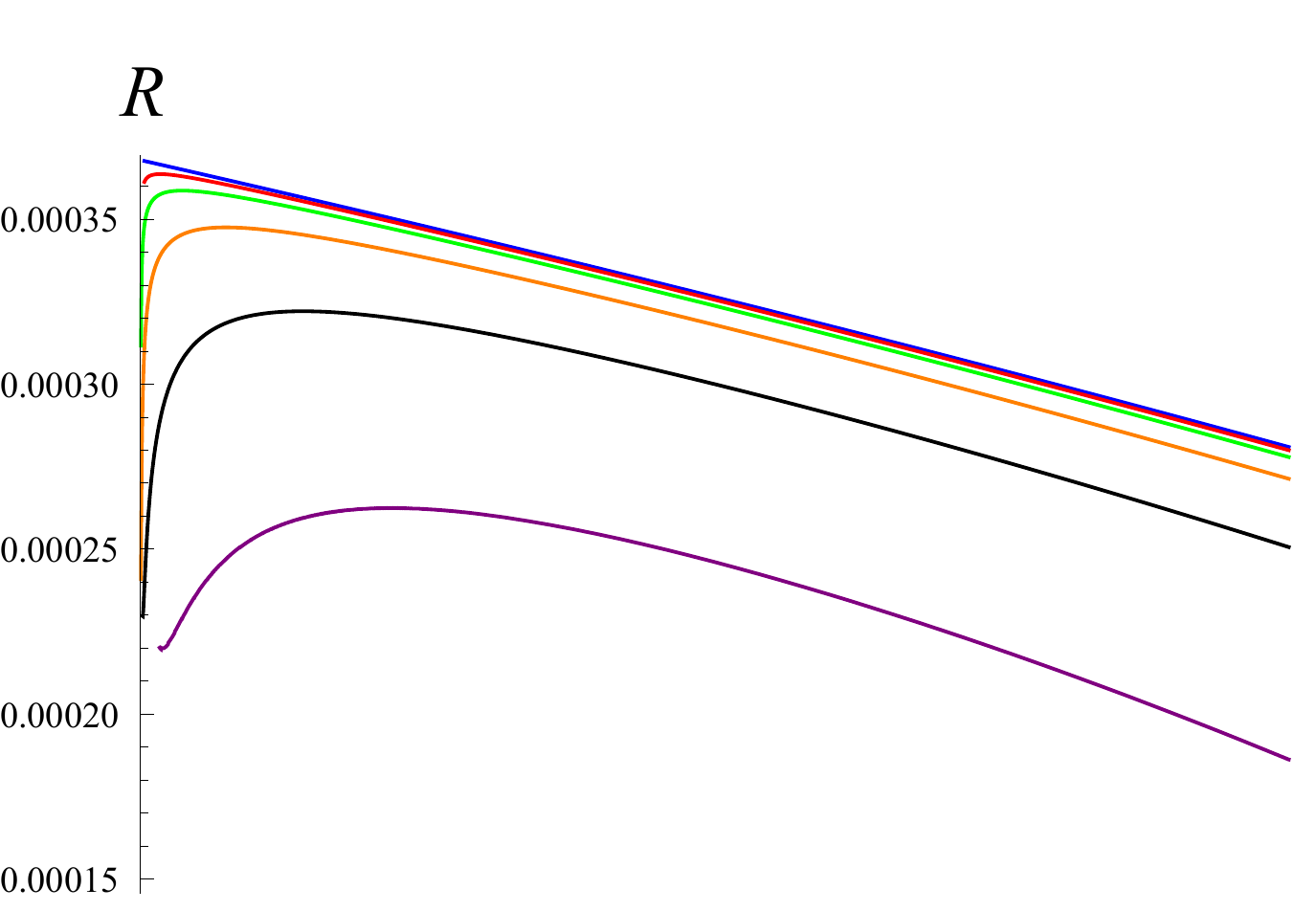}
\end{center}
\end{minipage}
\begin{minipage}{0.3\hsize}
\begin{center}
\scalebox{1.0}{\includegraphics[scale=0.6]{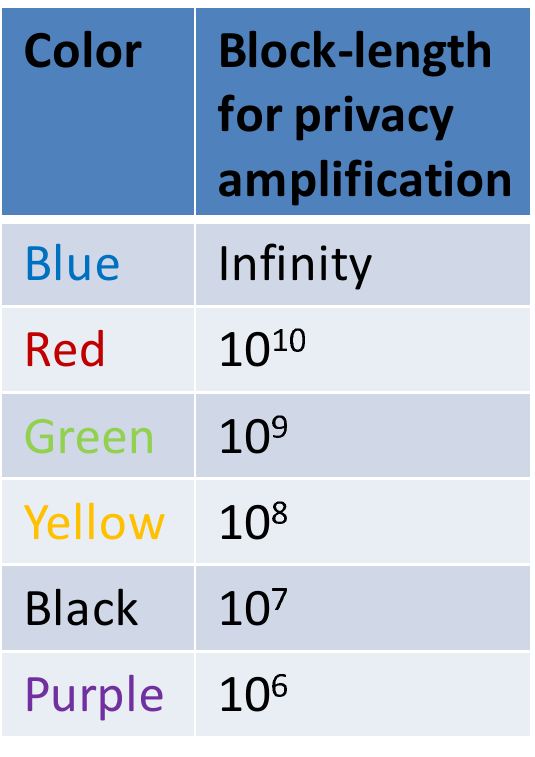}}
\end{center}
\end{minipage}
\end{tabular}
\caption{Key generation rate with weak coherent pulses:
We employ two intensities: signal intensity and decoy intensity.
Using the difference between detection rates of the pulses with two different intensities, 
we can estimate the fraction of multiple photons in the detected pulses.
Here, we set the signal intensity to be $1$.
This graph shows the key generation rate dependent on the decoy intensity.
This graph is based on the calculation formula given in \cite{H-N}.}
\Label{F5}
\end{figure}

\newpage

\subsection{History of developments of QKD}\Label{S5D}
Because the raw keys are not necessarily secure when the channel has noise or two photons are transmitted,
many studies have been done to find a way to guarantee security
when the communication device has such imperfections.
For this purpose, 
we need to consider a partial information leakage whose amount is bounded by the amount of the imperfection.
Shor and Preskill \cite{SP} and Mayers \cite{M01} showed that privacy amplification generates secure final keys
even when the channel has noise when the light source correctly generates a singlephoton.
Gottesman et al. \cite{GLLP} showed that these final keys can be secure
even when the light source occasionally generates multiple photons
if the fraction of multiple photon pulses is sufficiently is small.
The light source used in the actual quantum optical communication
is the weak coherent light, which probabilistically generates some multiple photon pulses, as shown in Fig. \ref{F9}.
Hence, this kind of extension had  been required for practical use.
Hwang \cite{decoy1} 
proposed an efficient method to estimate the fraction of multiple photon pulses, called the decoy method, in which the sender randomly chooses pulses with different intensities.

Until this stage, the studies of QKD were mainly done by individual researchers.
However, project style research is needed for a realization of QKD
because the required studies need more interactions between theorists and experimentalists.
A Japanese project, the ERATO Quantum Computation and Information Project,
tackled the problem of guaranteeing the security of a real QKD system.
Since this project contained an experimental group as well as theoretical groups,
this project naturally proceeded to a series of studies of QKD from a more practical viewpoint.
First, one project member, Hamada \cite{Ham2002,ou-Ham} studied the relation between the quantum error correcting code and the security of QKD more deeply.
Then, another project member, Wang \cite{decoy3,Wang05} extended the decoy method, which was developed independently
by a group at Toronto University \cite{decoy2,Ma05}.
Tsurumaru \cite{decoy4} and the author \cite{H1} have further extended the method. 
These extended decoy methods give a design for the choice of the intensity of transmitted pulses.
Further, jointly with the Japanese company NEC, the experimental group demonstrated
QKD with spools of standard telecom fiber over 100 km \cite{Tomi}. 

Here, we note that the theoretical results above assume the combination of error correction and privacy amplification
for an infinitely large block-length in steps (5) and (6).
They did not give a quantitative evaluation of the security with finite-block-length.
They also did not address
the construction of privacy amplification so 
these results are not sufficient for realization of a quantum key distribution system.
To resolve this issue, as a member of this project,
the author \cite{H-QKD} approximately evaluated the security  
with finite-block-length $n$
when the channel has noise and the light source correctly generates a single photon.
This idea has two key points.
The first contribution is the evaluation of information leakage via the phase error probability of virtual error correction in the phase basis, which 
is summarized as item (i).
This evaluation is based on the duality relation in quantum theory, which typically appears
in the relation between position and momentum.
The other contribution is the approximate evaluation of the phase error probability 
via the application of the central limit theorem, which is obtained by 
the combination of items (iii) and (iv).
This analysis is essentially equivalent to the derivation of 
the coefficient of the transmission rate up to the second-order $\frac{1}{\sqrt{n}}$. 
However, this analysis assumed a single-photon source.
Under this assumption, the author discussed the optimization for the ratio of check bits \cite{ORPB}.
Based on a strong request from the project leader of the ERATO project
and helpful suggestions by the experimental group, using the decoy method,
he extended a part of his analysis to the case when the light source sometimes generates multiple photons \cite{H-QKD2} by replacing item (iv) by (iv').
Based on this analytical result, 
the ERATO project made an experimental demonstration of QKD 
with weak coherent pulses on a real optical fiber, whose security is quantitatively guaranteed 
in the Gaussian approximation \cite{HHHTT}.
 
Another Japanese project of the 
National Institute of Information and Communication Technology (NICT)
has continuously made efforts toward a realization of QKD.
After the ERATO project, the author joined the NICT project from 2011 to 2016.
The NICT 
organized a project in Tokyo (Tokyo QKD Network) by connecting QKD devices operated by
NICT, NEC, Mitsubishi Electric, NTT, Toshiba Research Europe, ID Quantique, the Austrian Institute of Technology, the Institute of Quantum Optics and Quantum Information and the University of Vienna in 2010\cite{UQCC}.
Also, as a part of the NICT project, NEC developed a QKD system, as shown in Fig. \ref{F6B}, and performed a long-term evaluation experiment in 2015 \cite{Nec}.

After the above ERATO project, two main theoretical problems remained,
and their resolutions had been strongly required by the NICT project 
because they are linked to the security evaluation of these installed QKD systems.
The first one was the complete design of privacy amplification.
Indeed, in the above security analysis based on the phase error probability,
the range of possible random hash functions was not clarified. 
That is, only one example of a hash function was given in the paper \cite{H-QKD2},
and we had only a weaker version of item (ii) at that time.
To resolve this problem, as members of the NICT project,
Tsurumaru and the author clarified what kind of hash functions can be used to guarantee the security of a QKD system \cite{Tsuru},
which yields the current item (ii).
They introduced $\delta$-almost dual universal$_2$ hash functions, as explained in Section \ref{S5A2}.
In these studies, Tsurumaru taught the author 
the practical importance of the construction of hash functions from an industrial viewpoint
based on his experience obtained as a researcher at Mitsubishi Electric.

The second problem was to remove the Gaussian approximation in \cite{H-QKD}
from the finite-length analysis.
Usually, security analysis requires rigorous evaluation without approximation.
Hence, this requirement was essential for the security evaluation.
In Hayashi and Tsurumaru \cite{HT1}, we succeeded in removing this approximation and obtained 
a rigorous security analysis for the single-photon case.
Also, the paper \cite{HT1} clarified
the security criterion and simplified the derivation
in the discussion given in Subsection \ref{S5B}.
Based on a strong request by the NICT project,
the author extended the finite-length analysis to the case with multiple photons by employing the decoy method and performing a complicated statistical analysis \cite{H-N}.
The transmission rate in the typical case is shown in Fig. \ref{F5}.
This study clarified the requirements for physical devices to apply the obtained security formula.
In this study \cite{H-N}, 
the author also improved an existing decoy protocol. 
Under the improved protocol, he optimizes the choice of intensities \cite{ODI}.
Finally, we should remark that 
only such a mathematical analysis can guarantee the security of QKD.
This is quite similar to the situation that conventional security measures,
like RSA, can be guaranteed by mathematical analysis of the computational complexity \cite{RSA}.
In this way QKD is different from conventional communication technology.

Here, we should address the security analysis based on the leftover hash lemma \cite{BBCM,ILL} as another research stream of QKD.
This method came from cryptography theory 
and was started by the Renner group at the Swiss Federal Institute of Technology in Zurich (ETH) \cite{Renner1}.
The advantage of this method is the direct evaluation of information leakage without needing to evaluate the virtual phase error probability.
This method also enables a security analysis with finite-block-length \cite{TWGR}.
However, their finite-block-length analysis is looser than our analysis in Hayashi and Tsurumaru \cite{HT1} 
because their bound \cite{TWGR} cannot yield the second-order rate based on the central limit theorem
whereas it can be recovered from the bound in Hayashi and Tsurumaru \cite{HT1}. 
Further, while their method is potentially precise, 
it has very many parameters to be estimated in the finite-block-length analysis.
Although their method improves the asymptotic generation rate \cite{WMU},
the increase in the number of parameters to be estimated enlarges the error of channel estimation in the finite-length setting.
Hence, they need to decrease the number of parameters to be estimated.
In their finite-block-length analysis, they simplified their analysis so that 
only the virtual phase error probability has to be estimated.
This simplification improves the approach based on the leftover hash lemma because it gives a security evaluation based on the virtual phase error probability more directly.
However, this approach did not consider security with weak coherent pulses.
As another merit, the approach based on the leftover hash lemma later
influenced the security analysis in the classical setting \cite{Haya5,H-tight,H-ep,HW1,HT15}.

To discuss the future of QKD, 
we now describe other QKD projects.
Several projects were organized in Hefei in 2012 and in Jinan in 2013\cite{C-QKD}.
In 2013, a US company, Battelle, 
implemented a QKD system for commercial use in Ohio 
using a device from ID Quantique\cite{battelle}.
Battelle has a plan to establish a QKD system between Ohio and Washington, DC, over a distance of 700 km\cite{battelle}.
Also, in China, the Beijing-Shanghai project almost established a QKD system connecting Shanghai, Hefei, Jinan, and Beijing with over a distance of 2,000 km \cite{C-QKD}.
Indeed, these implemented QKD networks are composed of a collection of QKD communications over relatively short distance.
However, quite recently,
a Chinese group has succeeded in realizing a satellite for
quantum communications.
Since most of these developments are composed of networks of quantum communication channels,
it is necessary to develop theoretical results to exploit the properties of quantum networks for a QKD system.



\section{Second-order channel coding}\Label{S6}
Now, we return to classical channel coding with the memoryless condition.
In the channel coding, it is important to clarify 
the difference between the asymptotic transmission rate
and the actual optimal transmission rate dependent on the block-length, as shown in Fig. \ref{F11}.
This is because 
many researchers had mistakenly thought that 
the actual optimal transmission rate equals the asymptotic transmission rate for a long time.

\begin{figure}[htbp]
\begin{center}
\scalebox{0.5}{\includegraphics[scale=1]{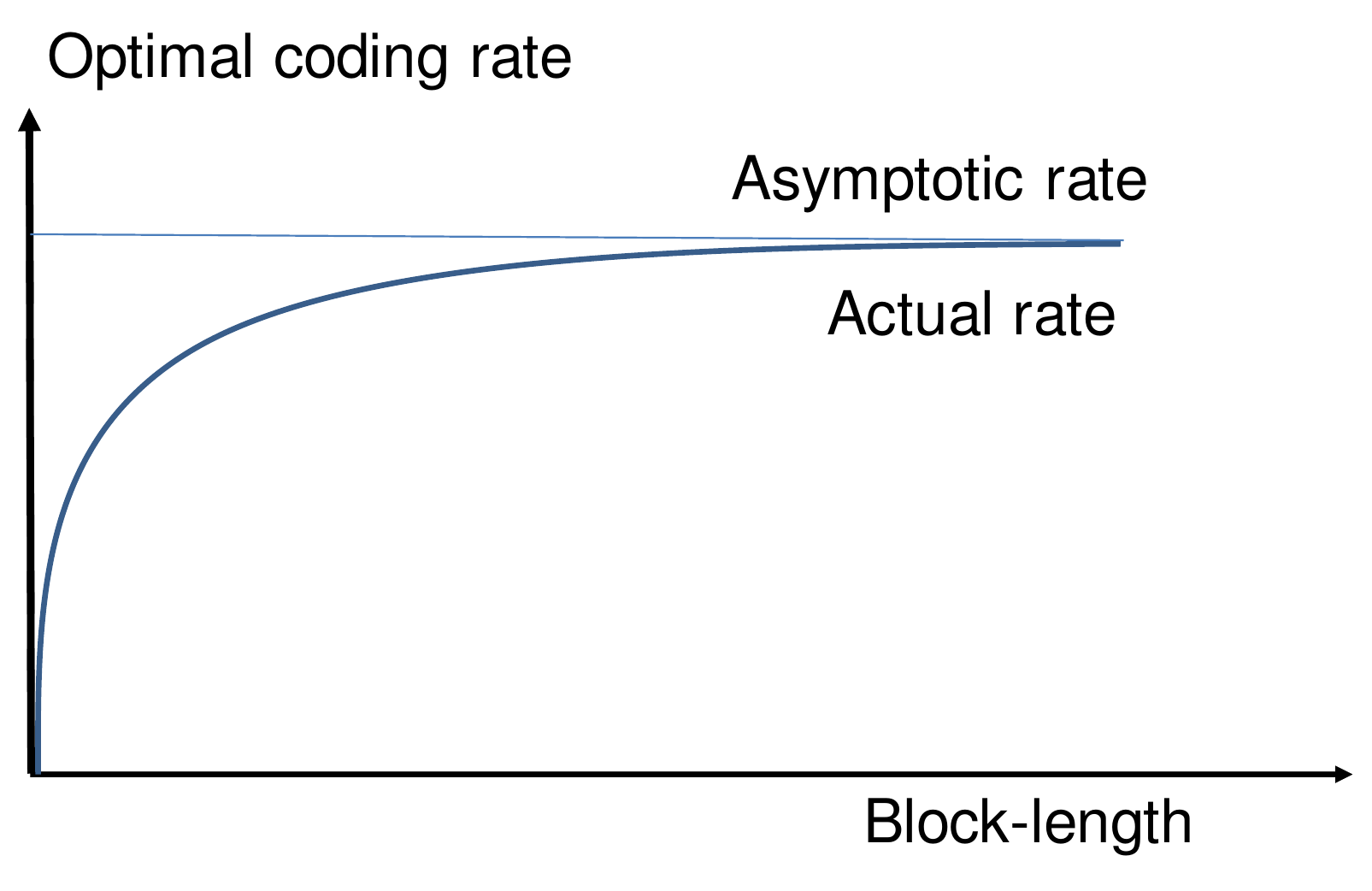}}
\end{center}
\caption{
Relation between the asymptotic transmission rate
and the actual transmission rate dependent on the block-length:
Usually, the actual transmission rate is smaller than the asymptotic key generation rate.
As the block-length increases, the actual transmission rate becomes closer to 
the asymptotic key generation rate.}
\Label{F11}
\end{figure}%

When the channel $P_{Y|X}$ is given as a binary additive noise subject to the distribution $P_Z$
and the channel $P_{Y^n|X^n}$ is the product distribution of the channel $P_{Y|X}$,
the simple combination of \eqref{8-31-ab} and \eqref{12-9-1} yields 
the  asymptotic expansion of
$M(\epsilon| P_{Y^n|X^n})$: 
\begin{align}
M(\epsilon| P_{Y^n|X^n})
= n (\log 2-H(P_Z))
+ \sqrt{n}\sqrt{V(P_Z)}\Phi^{-1}(\epsilon)+ o(\sqrt{n})
\end{align}
because Eq. \eqref{8-31-ab} does not contain $\sup_{P_{X_n}}$ like \eqref{8-31-a}.
In the general case, 
using the formulas \eqref{8-31-5} or \eqref{8-31-a} with order $\sqrt{n}$,
we can derive the $\ge $ part of the following inequality.
\begin{align}
M(\epsilon| P_{Y^n|X^n})
=
\left\{
\begin{array}{ll}
n \max_{P_X}I(P_X,P_{Y|X})
+ \sqrt{n}\sqrt{V_-(P_{Y|X})}\Phi^{-1}(\epsilon)+ o(\sqrt{n})
& \hbox{ if } \epsilon <\frac{1}{2} \\
n \max_{P_X}I(P_X,P_{Y|X})
+ \sqrt{n}\sqrt{V_+(P_{Y|X})}\Phi^{-1}(\epsilon)+ o(\sqrt{n})
& \hbox{ if } \epsilon \ge \frac{1}{2} ,
\end{array}
\right.
\Label{9-1-4}
\end{align}
where $V(P_{Y|X})$ is defined as
\begin{align}
V_+(P_{Y|X}) &:=
\max_{P_X} \sum_{x}P_X(x) \sum_y P_{Y|X}(y|x)
\Big(\log \frac{P_{Y|X}(y|x)}{P_Y(y)}- D(P_{Y|X=x}\|P_Y)\Big)^2 
\Label{9-1-5}\\
V_-(P_{Y|X}) &:=
\min_{P_X} \sum_{x}P_X(x) \sum_y P_{Y|X}(y|x)
\Big(\log \frac{P_{Y|X}(y|x)}{P_Y(y)}- D(P_{Y|X=x}\|P_Y)\Big)^2 ,
\Label{9-1-6}
\end{align}
and the minimum and maximum are taken over the $P_X$ satisfying 
$I(P_X,P_{Y|X})= \max_{Q} I(Q,P_{Y|X})$.
However, it is difficult to derive the $\le$ part of inequality \eqref{9-1-4}
by using \eqref{8-31-a} due to the maximization $\sup_{P_{X_n}}$.

To resolve this problem,
we choose $P_{X}$ as the distribution realizing 
the minimum in \eqref{9-1-6}
or the maximum in \eqref{9-1-5}
and substitute 
$P_{X}^n$ into $Q_n$ in the formula \eqref{8-31-b}.
Then, we can derive the $\le$ part of the inequality \eqref{9-1-4}.
Although this expansion was firstly derived by Strassen \cite{strassen} in 1962,
this derivation is much simpler, which shows 
the effectiveness of the method of information spectrum.

The author applied this method to an additive white Gaussian noise channel 
and succeeded in deriving the second-order coefficient of its transmission rate, which had been unknown until that time;
this was published in 2009 \cite{Hay1}.
In fact, he obtained only a rederivation of Strassen's result.
When he presented this result in a domestic meeting \cite{second-D},
Uyematsu pointed out Strassen's result.
To go beyond Strassen's result, he applied this idea to the additive white Gaussian noise channel, and obtained the following expansion,
which appears as a typical situation in wireless communication.
\begin{align}
\log M(\epsilon|S,N)
=
\frac{n}{2}\log \Big(1+\frac{S}{N}\Big) 
+ 
\sqrt{n}\frac{\frac{S^2}{N^2}+\frac{2S}{N}}{2(1+\frac{S}{N})}
\Phi^{-1}(\epsilon)+ o(\sqrt{n}),
\Label{9-1-8}
\end{align}
where
$M(\epsilon|S,N)$ is the maximum size of transmission when the variance of the Gaussian noise is $N$ and the power constraint is $S$.

In fact, a group in Princeton University, mainly Verd\'{u} and Polyanskiy,
tackled this problem independently.
In their papers \cite{Pol,Pol2},
they considered the relation between channel coding and simple statistical hypothesis testing,
and independently derived two relations, the dependence test bound and the meta converse inequality,
which are the same as in the classical special case considered in the author and Nagaoka\cite{Hay-Nag} and Nagaoka \cite{Naga-EQIS}.  
Since their results \cite{Pol} are limited to the classical case, 
the applicable region of their results is narrower than that of the preceding results in \cite{Hay-Nag,Naga-EQIS}.
Then, Verd\'{u} and Polyanskiy rederived Strassen's result, without use of the method of information spectrum, by the direct evaluation of these two bounds.
They also independently derived the second-order coefficient of the optimal transmission rate for the additive white Gaussian noise channel in 2010 \cite{Pol}.
Since the result by this group at Princeton had a large impact in the 
information theory community at that time,
their paper received the best paper award of IEEE Information theory society in 2011 
jointly with the preceding paper by the author \cite{Hay1}.

As explained above, the Japanese group obtained some of the same results 
several years before the Princeton group but had much weaker publicity than the Princeton group.
Thus, the Princeton group met the demand in the information theory community,
and they presented their results very effectively.
In particular, since their research activity was limited to the information theory community,
their audiences were suitably concentrated so that they could create a scientific boom in this direction.
In contrast to the Princeton group, the Japanese group studied the same topic far from the demand of the community because their study originated in quantum information theory.
In particular, their research funds were intended for the study of quantum information so they had to present their work to quantum information audiences who are less interested in their results.
Also, because their work was across too wide a research area to explain their results effectively,
they could not devote sufficient efforts to explain their results to the proper audiences at that time.
Hence, their papers attracted less attention.
For example, few Japanese researchers knew the paper \cite{Hay1} 
when it received the IEEE award in 2011.
After this award, this research direction became much more popular and was applied to very many topics in information theory \cite{PPV2,PPV3,TK,SKT15,KV,HW13,HW14c,HW1,YHN}.
In particular, the third-order analysis has been applied to channel coding \cite{TT13}.
These activities were reviewed in a recent book \cite{T14}.

Although this research direction is attracting much attention,
we need to be careful about evaluating its practical impact.
These studies consider finite-block-length analysis for the optimal rate with respect to all codes including those with too high a calculation complexity to implement.
Hence, the obtained rate cannot necessarily be realized with implementable codes.
To resolve this issue,
we need to discuss the optimal rate among codes whose calculation complexity is not so high.
Because no existing study discusses this type of finite-block-length analysis,
such a study is strongly recommend for the future.
Also, a realistic system is not necessarily memoryless;
so, we need to discuss memory effects.
To resolve this issue, jointly with Watanabe,
the author extended this work to channels with additive Markovian noise,
which covers the case when Markovian memory exists in the channel \cite{HW13}.
While this model covers many types of realistic channel,
it is not trivial to apply the results in \cite{HW13} to the realistic case of wireless communication
because it is complicated to address the effect of fading in the coefficients.
This is an interesting future problem.

After this breakthrough, the
Princeton group extended their idea to many topics in channel coding and data compression \cite{PPV2,PPV3,KV,KV14}.
On the other hand,  
in addition to the above Markovian extension,
the author, jointly with Tomamichel, extended this work to the quantum system \cite{T-M}, providing a unified framework for the second-order theory in the quantum system
for data compression with side information, secure uniform random number generation,
and simple hypothesis testing.
At the same time, Li \cite{KL} directly derived the second-order analysis for simple statistical hypothesis testing in the quantum case.
However, the second-order theory for simple statistical hypothesis testing 
has less meaning in itself;
it is more meaningful in relation to other topics in information theory.

\section{Extension to physical layer security}\Label{S7}
\subsection{Wire-tap channel and its variants}
The quantum cryptography explained above offers secure key distribution based on physical laws.
The classical counterpart of quantum cryptography is physical layer security,
which offers information theoretical security based on several physical assumptions from classical mechanics. 
As its typical mode, Wyner \cite{Wyner} formulated the wire-tap channel model,
which was more deeply investigated by Csisz\'{a}r and K\"{o}ner\cite{CK79}.
This model assumes two channels, as shown in Fig. \ref{F10x}: 
a channel $P_{Y|X}$ from the authorized sender (Alice) to the authorized receiver (Bob)
and a channel $P_{Z|X}$ from the authorized sender to the eavesdropper (Eve).
When the original signal of Alice has stronger correlation 
with the received signal than that with Eve, that is,
a suitable input distribution $P_X$ satisfies
the condition $I(P_X,P_{Y|X}) > I(P_X,P_{Z|X})$,
the authorized users can communicate without any information leakage
by using a suitable code.
More precisely, secure communication is available
if and only if there exists a suitable joint distribution $P_{VX}$ between the input system ${\cal X}$ and another system ${\cal V}$
such that the condition $I(P_V,P_{Y|V}) > I(P_V,P_{Z|V})$ holds,
where $P_{Y|V}(y|v):=\sum_{x\in {\cal X}}P_{Y|X}(y|x)P_{X|V}(x|v)$
and $P_{Z|V}$ is defined in the same way.

\begin{figure}[htbp]
\begin{center}
\scalebox{0.7}{\includegraphics[scale=1]{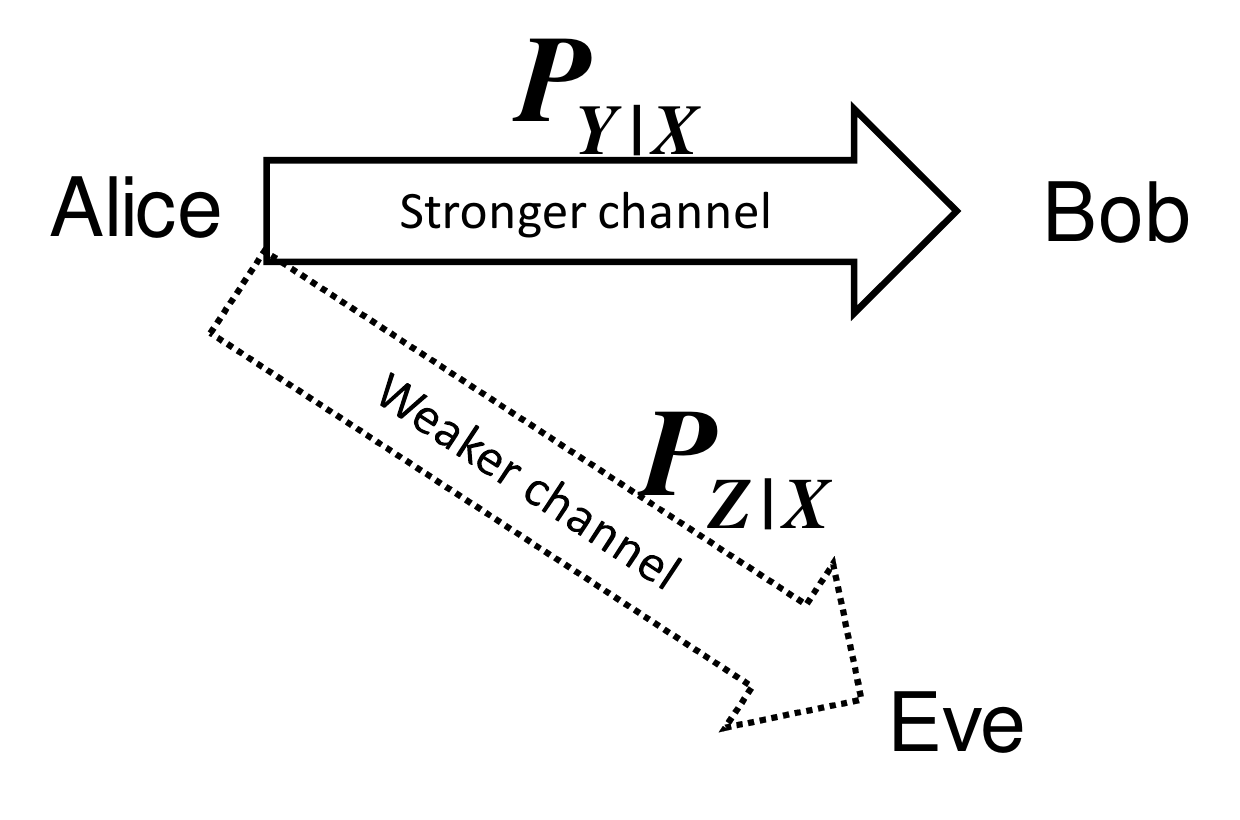}}
\end{center}
\caption{Wire-tap channel model. Eve is assumed to have a weaker connection to Alice than Bob does.} 
\Label{F10x}
\end{figure}%

Although we often assume that the channel is stationary and memoryless,
the general setting can be discussed by using information spectrum  \cite{Haya6}. 
This paper explicitly pointed out that there is a relation between the wire-tap channel and the channel resolvability discussed in Section \ref{S3}.
This idea has been employed in many subsequent studies \cite{BL,HG,HES}.
Watanabe and the author \cite{WH} discussed the second-order asymptotic for the channel resolvability.
Also, extending the idea of the meta converse inequality to the wire-tap channel,
Tyagi, Watanabe, and the author showed a relation between the wire-tap channel and hypothesis testing\cite{HTW1}.
Based on these results, Yang et al. \cite{YSP} investigated finite-block-length bounds for wiretap channels without Gaussian approximation.
Also, taking into account the construction complexity,
the author and Matsumoto \cite[Section XI]{HM} proposed another type of finite-length analysis for wire-tap channels. 
Its quantum extension has also been discussed \cite{Deve,H-wire}.
However, in the wire-tap channel model, we need to assume that Alice and Bob know the channel $P_{Z|X}$ to Eve.
Hence, 
although it is a typical model for information theoretic security,
this model is somewhat unrealistic because Alice and Bob cannot identify Eve's behavior.
That is, it is assumed that Eve has weaker connection to Alice than Bob does, as shown in Fig. \ref{F10x}.
So, it is quite hard to find a realistic situation where the original wire-tap channel model is applicable.

Fortunately, this model has more realistic derivatives:
one is secret sharing\cite{Bla,Sha}, and another is secure network coding\cite{bhattad05,cai11survey,cai02b,cai07securecondition,caiyeung11,HY}.
In secret sharing, 
there is one sender, Alice, and $k$ receivers, Bob$1$, $\ldots$, Bob$k$.
Alice splits her information into $k$ parts, and sends them to the respective Bobs
such that 
a subset of Bobs cannot recover the original message.
For example, assume that there are two Bobs, $X_1$ is the original message and $X_2$ is an independent uniform random number.
If Alice sends the exclusive or of $X_1$ and $X_2$ to Bob$1$
and sends $X_2$ to Bob$2$,
neither Bob can recover the original message.
When both Bobs cooperate, however, they can recover it.
In the general case, 
for any given numbers $k_1<k_2 < k$, we manage our code such that
any set of $k_1$ Bobs cannot recover the original message but any set of $k_2$ Bobs can \cite{Yamamoto}.

Secure network coding is a more difficult task.
In secure network coding, Alice sends her information to the receiver via a network, and
the information is transmitted to the receiver via intermediate links. 
That is, each intermediate link transfers a part of the information.
Secure network coding is a method to guarantee security when some of the intermediate links are eavesdropped by Eve.
Such a method can be realized by applying the wire-tap channel to the case when Eve obtains the information from some of intermediate links\cite{bhattad05,cai11survey,cai02b,cai07securecondition,caiyeung11,HY}.
When each intermediate link has the same amount of information,
the required task can be regarded as a special case of secret sharing.

However, this method depends on the structure of the network,
and it is quite difficult for Alice to know this structure.
Hence, it is necessary to develop a coding method that does not depend on the structure of the network.
Such a coding is called universal secure network coding,
and has been developed by several researchers\cite{silva08,silva09,NT,KMU,Kurosawa}.
These studies assume only that the information processes on each node are linear
and the structure of network does not change during transmission.
In particular, the security evaluation can be made even with finite-block-length codes \cite{NT,KMU,Kurosawa}.
Since it is quite hard to tap all of the links,
this kind of security is sufficiently useful for practical use by ordinary people 
based on the cost-benefit analysis of performance.
To understand the naturalness of this kind of assumption,
let us consider the daily-life case in which an important message is sent by dividing it into two e-mails,
the first of which contains the message encrypted by a secure key, and the second one contains the secure key.
This situation assumes that only one of two links is eavesdropped.

\subsection{Secure key distillation}
As another type of information theoretical security,
Ahlswede and Csisz\'{a}r\cite{Ahlswede} and Maurer\cite{Mau93} proposed 
secure key distillation. 
Assume that two authorized users, Alice and Bob, have random variables $X$ and $Y$,
and the eavesdropper, Eve, has another random variable $Z$.
When the mutual information $I(X;Y)$ between Alice and Bob is larger than 
the mutual information $I(X;Z)$ or $I(Y;Z)$ between one authorized user and Eve,
and
when their information is given as the $n$-fold iid distribution of a given joint distribution $P_{XYZ}$,
Alice and Bob can extract secure final keys.

Recently, secure key distillation has been developed in a more practical way
by importing the methods developed for or motivated by quantum cryptography 
\cite{Haya5,H-tight,H-ep,HW1,HT15,Watanabe,BTH}.
In particular, its finite-block-length analysis
has been much developed, including the Markovian case, when Alice's random variable agrees with Bob's random variable
\cite{Haya5,H-tight,H-ep,HW1,HT15}.
Such a analysis has been extended to a more general case in which
Alice's random variable does not necessarily agree with Bob's random variable \cite{HTW2}.
Although some of the random hash functions were originally constructed for quantum cryptography, 
they can be used for privacy amplification even in secure key distillation \cite{Carter,WC81,HT2}.
Hence, under several natural assumptions for secure key distillation,
it is possible to precisely evaluate the security based on finite-block-length analysis. 

We assume that $X$ is a binary information, and 
all information is given as the $n$-fold iid distribution of a given joint distribution $P_{XYZ}$.
In this case, the protocol is essentially given by steps (5) and (6) of QKD,
where the code $C$, its dimension $k$, and the sacrifice bit length $\bar{k}$ are determined a priori according to the joint distribution $P_{XYZ}$.
Now, we denote the information exchanged via the public channel by $u$
and its distribution by $P_{{\rm pub}}$.
The security is evaluated by the criterion;
\begin{align}
\gamma(C,\{f_r\}):=
\frac{1}{2}
\sum_{u}\sum_{r} P_R(r)
P_{{\rm pub}}(u) \sum_{x\in \bF_2^{k-\bar{k}}} \sum_{z \in {\cal Z}^n}
| P_{f_{r}(X^n) Z^n|U}(x,z|u) -  
P_{ \bF_2^{k-\bar{k}},\mix}(x)P_{Z^n|U}(z|u)|
\Label{9-2-1},
\end{align}
where $P_R$ is the distribution of the random variable $R$ used to choose our hash function $f_R$. 
To evaluate this criterion, we introduce the conditional R\'{e}nyi entropy
\footnote{Indeed, two kinds of conditional R\'{e}nyi entropy are known.
This type is often called the Gallager \cite{Gallager} or Arimoto type\cite{Arimoto}.}
\begin{align}
H_{1+s}(X|Z|P_{XZ}):= -\frac{1+s}{s} \log 
\Big(\sum_{z\in {\cal Z}}P_{Z}(z)
\Big(\sum_{x\in {\cal X}}P_{X|Z}(x|z)^{1+s}
\Big)^{\frac{1}{1+s}}
\Big).
\end{align}
Then, the criterion is evaluated as (\cite[(54) and Lemma 22]{H-ep} and \cite[(21)]{H-q-secure}\footnote{For its detail derivation, see \cite[Section V-D]{H-wireless}.})
\begin{align}
\gamma(C,\{f_r\})\le 
(1+\frac{\sqrt{\delta}}{2})
\min_{s \in [0,1]}
e^{\frac{s}{1+s} (n \log 2-\bar{k} -n H_{1+s}(X|Z|P_{XZ}) )}.\Label{9-2-2}
\end{align}
Its quantum extension has also been discussed in \cite{DW,H-q-secure}.

Here, we should remark that the evaluation \eqref{9-2-2} can be realized by 
a random hash function with small calculation complexity.
This is because 
the inequality holds for an arbitrary linear code and an arbitrary $\delta$-almost dual universal$_2$ hash function.
Since the paper \cite{HT2} proposed several efficient $\delta$-almost dual universal$_2$ hash functions,
the bound has operational meaning even when we take into account the calculation complexity for its construction.

So, one might consider that secure key distillation is the same as QKD.
However, QKD is different from secure key distillation even with the quantum extension
due to the following points.
The advantage of QKD is that it does not assume anything except for the basic laws of quantum 
theory.
Hence, QKD usually does not allow us to make any additional assumptions,
in particular, the iid assumption.
In contrast, in secure key generation, we often make the iid assumption.
As another difference, we are assumed to know the joint distribution or the density matrix on the whole system in secure key distillation
whereas we need to estimate only the density matrix on the whole system in QKD.

The finite-block-length analysis of secure key distillation
is different from that for channel coding in the following point.
The obtained finite-block-length analysis for channel coding discusses only the optimal performance among all codes, including impractical codes whose calculation complexity is too high.
However, in the finite-block-length analysis for physical layer security,
the obtained bound can be attained by a practical protocol whose calculation complexity is linear in the block-length.

\subsection{Application to wireless communication}
Recently, along with the growing use of wireless communication,
secure wireless communication has been very actively studied \cite{YMRSTM,WS,PJCG,CJ,Trappe,Zeng,WX}.
Physical layer security has been considered as a good candidate for secure wireless communication \cite{BBRM,YPS}.
Typically,
we assume the quasi-static condition, which allows us to assume
the memoryless condition in one coding block-length.
Even with this condition, a simple application of the wire-tap channel cannot guarantee secure communication
when Eve set up her antenna between Alice and Bob.
However, when the noise in Bob's output signal is independent of the noise in Eve's output signal, 
the mutual information between Alice and Bob is larger than that between Eve and Bob even in this situation.
In this case, when they apply secure distillation in the opposite way after the initial wireless communication from Alice to Bob,
they can generate secure keys.
The assumption of the independence between Bob's and Eve's outputs is too strong and unrealistic for a practical use
because there is a possibility of interference between the two channels.
Hence, a more realistic assumption is needed.

\begin{figure}[htbp]
\begin{center}
\scalebox{0.7}{\includegraphics[scale=1]{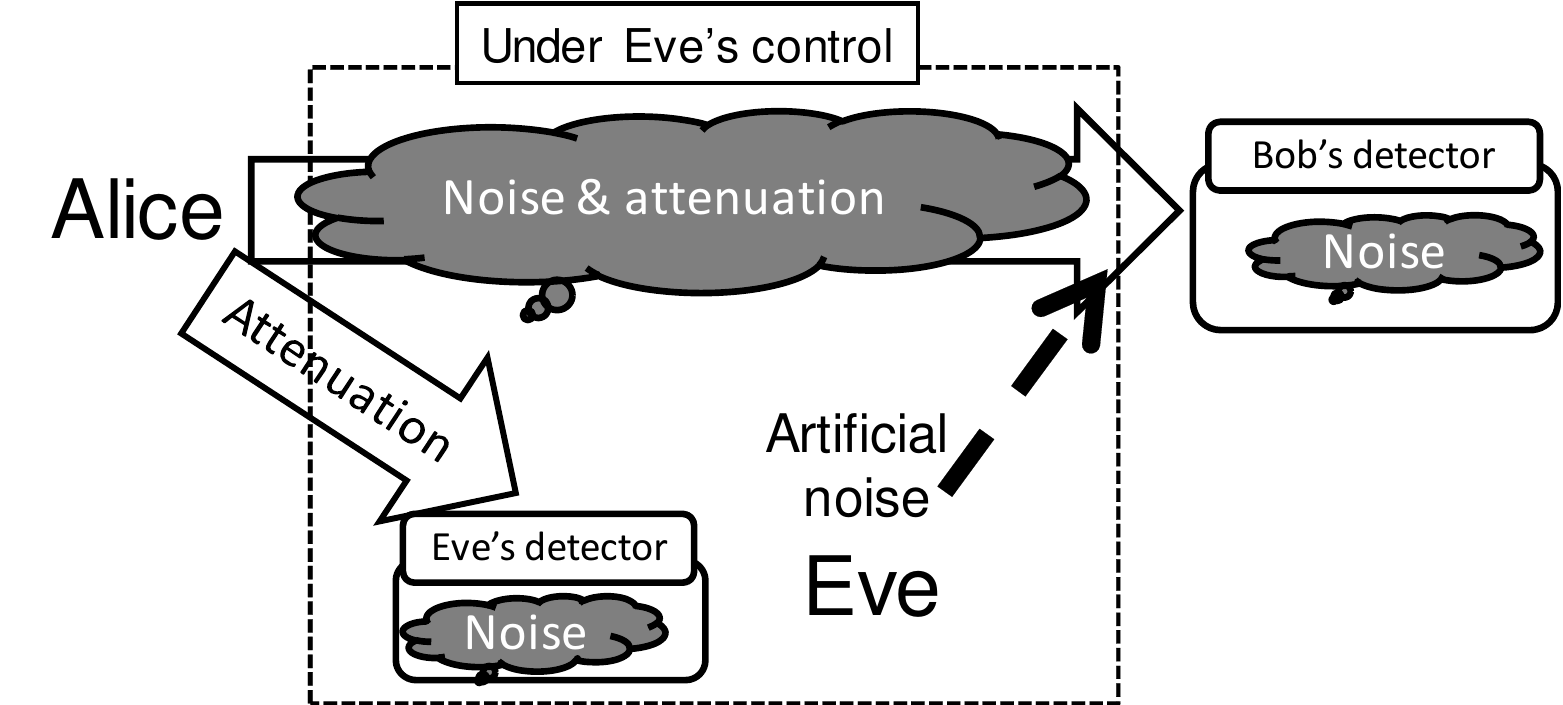}}
\end{center}
\caption{Model of Eve's attack for secure wireless communication. Eve can inject artificial noise into Bob's observation.
It is also assumed that Eve has noise in her detector like Bob.
It is natural to assume that these detector noises are independent of other random variables.}
\Label{F10}
\end{figure}%

To resolve this problem, 
the author had the following idea based on the experience of interactions with experimentalists studying QKD.
It is natural to assume that 
the noises generated inside each detector are independent and Gaussian,
and only the noise generated outside the detector is correlated to Eve's output.
In this situation, even when all of the intermediate space between Alice and Bob is under the control of Eve 
and Eve injects artificial noise into Bob's observation, as in Fig. \ref{F10}, 
when the noise is sufficiently small, 
the author showed that Alice and Bob can still generate secure keys \cite{H-wireless}.
Here, after the communication via the noisy wireless channel, 
Alice and Bob need to estimate the noise by random sampling.
Once the random sampling guarantees that the noise is sufficiently small, they apply the secure key generation protocol.
This is a protocol to generate secure keys between Alice and Bob under reasonable assumptions for secure wireless communication.
Although the paper \cite{H-wireless} gives such a protocol with a small calculation complexity for construction,
the real performance of this protocol has not been studied in real situations.
A future task is to estimate the performance of the proposed method in a realistic situation by taking into account several complicated realistic effects, including fading.

Here, we summarize the advantages over modern cryptography based on computation complexity\cite{Trappe}.
When cryptography based on computation complexity is broken by a computer,
any information transmitted with this cryptography can be eavesdropped by using that computer.
To break physical layer security of the above type,
Eve has to set up 
an antenna for each communication.
Furthermore, each antenna must be very expensive because 
it must break the above assumption.
Maybe, it is not impossible to break a limited number of specific communications 
for a very limited number of persons.
However, due to the cost constraint, it is impossible to eavesdrop on all communications
in a realistic situation.
In this way, physical layer security offers a different type of security from computational security.


\section{Conclusions and future prospects}
In this review article, we have discussed developments of finite-block-length theory in 
classical and quantum information theory:
classical and quantum channel coding, data compression,
(secure) random number generation,
quantum cryptography,
and physical layer security.
These subareas have been developed with strong interactions with each other
in unexpected ways.

The required future studies for channel coding and data compression 
are completely different from those needed for security topics.
In the former topics,
existing finite-block-length theory discusses only the minimum error among all codes 
without considering the calculation complexity of its construction.
Hence, for practical use, 
we need a finite-block-length theory for realizable codes whose 
construction has less calculation complexity.
Such finite-block-length theory is urgently required. 
Fortunately, the latest results obtained for these two topics 
\cite{HW13,HW14c} cover the case when a Markovian memory effect exists.
However, their applications to a realistic situation have not been sufficiently studied, and
such practical applications are interesting open problems.

In contrast, in the latter topics, the established finite-block-length theory already 
takes into account the calculation complexity of its construction;
hence, it is more practical.
However, these types of security protocols have not been realized for the following reasons.
In the case of quantum cryptography, we need more development on the device side.
Also, to realize secure communication for distances over 2000 km, we might need another type of information-scientific combinatorics.
In the case of physical layer security, 
we need more studies to fill the gap between information theoretical security analysis
and device development. 
There has recently been one such study \cite{H-wireless}.

Furthermore, the idea of finite-block-length theory is fundamental and can be extended to areas beyond information theory.
For example, it has been applied to  
a statistical mechanical rederivation of thermodynamics \cite{TH14,HT15b},
the conversion of entangled states \cite{WK13,WK13b,WK14,IWK15}, and
the analysis of the gap between two classes of local operations \cite{HO}.
Therefore, we can expect more applications of finite-block-length theory to other areas.

\section*{Acknowledgments}

The works reported here were supported in part by 
a MEXT Grant-in-Aid for Scientific Research (A) No. 23246071,
a JSPS Grant-in-Aid for Young Scientists (A) No. 20686026,
a JSPS Grant-in-Aid for Young Scientists (B) No. 14750330,
a JSPS Grant-in-Aid for Scientific Research on Priority Area ``Deepening
and Expansion of Statistical Mechanical Informatics (DEX-SMI)'' No. 18079014,
ERATO(-SORST) Quantum Computation and Information Project of the
Japan Science and Technology Agency (JST), 
the Brain Science Institute of RIKEN,
the National Institute of Information and Communication Technology (NICT), Japan,
the Okawa Research Grant,
and the Kayamori Foundation of Informational Science Advancement.
The author is grateful to 
Professor Akihisa Tomita, who is an expert on the physical implementation of QKD systems,
for discussing the physical model for a real QKD system.
He is also thankful to  
Dr. Toyohiro Tsurumaru and Dr. Kiyoshi Tamaki, who are working on QKD from an industrial perspective,
for discussing the physical assumptions for the decoy model.
He is also grateful to 
Professor \'{A}ngeles Vazquez-Castro,
Professor Hideichi Sasaoka, and Professor Hisato Iwai, 
who are experts in wireless communication,
for discussing the validity of the model of the paper \cite{H-wireless} for 
secure wireless communication.
He is also thankful to Dr. Ken-ichiro Yoshino in NEC for providing the picture of the QKD device (Fig. \ref{F6B}).


\end{document}